\documentclass[12pt]{article}
\usepackage[latin1]{inputenc}

\usepackage{amsmath}
\usepackage{color}
\usepackage{amsfonts}
\usepackage{amssymb}
\usepackage{graphicx}
\usepackage{geometry}
\usepackage{amssymb,epsfig,subfigure}
\usepackage{hyperref}
\usepackage{comment}
\usepackage{tabularx}
\usepackage{bm}
\usepackage{euscript}
\usepackage{graphicx}
\usepackage{color}
\usepackage{amsfonts}
\usepackage{exscale}
\usepackage{amsbsy}
\usepackage{subfigure}
\usepackage{textcomp}
\usepackage{comment}
\usepackage{hyperref}
\usepackage{slashed}
\usepackage{authblk}

\usepackage{tabularx}
\usepackage{euscript}
\usepackage{graphicx}
\usepackage{color}
\usepackage{amsfonts}
\usepackage{exscale}
\usepackage{amsbsy}
\usepackage{subfigure}
\usepackage{textcomp}
\usepackage{comment}
\usepackage{hyperref}
\usepackage{bm}

\newcommand{\be}{\begin{equation}}
\newcommand{\ee}{\end{equation}}
\newcommand{\bea}{\begin{eqnarray}}
\newcommand{\eea}{\end{eqnarray}}

\newcommand{\pd}{\partial}

\newcommand{\matleft}{\left(\begin{array}}
\newcommand{\matright}{\end{array}\right)}

\newcommand{\sgn}{\operatorname{sgn}}

\usepackage[numbers,sort&compress]{natbib}

\def\ba#1\ea{\begin{align}#1\end{align}}
\def\bg#1\eg{\begin{gather}#1\end{gather}}
\def\bm#1\em{\begin{multline}#1\end{multline}}
\def\bmd#1\emd{\begin{multlined}#1\end{multlined}}

\def\simge{%  ``greater than about'' symbol
    \mathrel{\rlap{\raise 0.511ex 
        \hbox{$>$}}{\lower 0.511ex \hbox{$\sim$}}}}

\def\simle{%  ``less than about'' symbol
    \mathrel{\rlap{\raise 0.511ex 
        \hbox{$<$}}{\lower 0.511ex \hbox{$\sim$}}}}

\makeatletter
\renewcommand\section{\@startsection {section}{1}{\z@}%
                                 {-3.5ex \@plus -1ex \@minus -.2ex}%nn
                                   {2.3ex \@plus.2ex}%
                                   {\normalfont\large\bfseries}}
\renewcommand\subsection{\@startsection{subsection}{2}{\z@}%
                                   {-3.25ex\@plus -1ex \@minus -.2ex}%
                                     {1.5ex \@plus .2ex}%
                                     {\normalfont\bfseries}}
\renewcommand\subsubsection{\@startsection{subsubsection}{3}{\z@}%
                                   {-3.25ex\@plus -1ex \@minus -.2ex}%
                                     {1.5ex \@plus .2ex}%
                                     {\normalfont\itshape}}
\makeatother

\def\pplogo{\vbox{\kern-\headheight\kern -29pt
\halign{##&##\hfil\cr&{\ppnumber}\cr\rule{0pt}{2.5ex}&\ppdate\cr}}}
\makeatletter
\def\ps@firstpage{\ps@empty \def\@oddhead{\hss\pplogo}%
  \let\@evenhead\@oddhead % in case an article starts on a left-hand page
}%      The only change in \maketitle is \thispagestyle{firstpage} instead of 
\def\maketitle{\par
 \begingroup
 \def\thefootnote{\fnsymbol{footnote}}
 \def\@makefnmark{\hbox{$^{\@thefnmark}$\hss}}
 \if@twocolumn
 \twocolumn[\@maketitle]
 \else \newpage
 \global\@topnum\z@ \@maketitle \fi\thispagestyle{firstpage}\@thanks
 \endgroup
 \setcounter{footnote}{0}
 \let\maketitle\relax
 \let\@maketitle\relax
 \gdef\@thanks{}\gdef\@author{}\gdef\@title{}\let\thanks\relax}
\makeatother

\numberwithin{equation}{section}

\begin{document}

\setcounter{page}0
\def\ppnumber{\vbox{\baselineskip14pt
%\hbox{hep-th/0000000}
}}

\def\ppdate{%\footnotesize{SU/ITP-14/XX}
} \date{\today}

\title{\LARGE \bf Loop Models, Modular Invariance, and \\ Three Dimensional Bosonization}
\author{Hart Goldman and Eduardo Fradkin}
\affil{\it Department of Physics and Institute for Condensed Matter Theory,\\  \it University of Illinois at Urbana-Champaign, \\  \it 1110 West Green Street, Urbana, Illinois 61801-3080, USA}

\bigskip

\maketitle

\begin{abstract}
We consider a family of quantum loop models in 2+1 spacetime dimensions with marginally long-ranged and statistical interactions mediated by a $U(1)$ gauge field, both purely in 2+1 dimensions and on a surface in a 3+1 dimensional bulk system. %Such interactions are mediated by a U$(1)$ gauge field. 
In the absence of fractional spin, these theories have been shown to be self-dual under particle-vortex duality and shifts of the statistical angle of the loops by $2\pi$, which form a subgroup of the modular group, PSL$(2,\mathbb{Z})$. We show that careful consideration of fractional spin in these theories completely breaks their statistical periodicity and describe how this occurs, resolving %an apparent inconsistency 
a disagreement with the conformal field theories they appear to approach at criticality. %As a result, we argue that the action of PSL(2,$\mathbb{Z}$) on the loop models has a natural relationship with the PSL(2,$\mathbb{Z}$) which organizes the recent web of 2+1 dimensional field theory dualities. 
We show explicitly that incorporation of fractional spin leads to loop model dualities which parallel %are consistent with members of 
the recent web of 2+1 dimensional field theory dualities,
%with the duality web
 providing a nontrivial check on its validity.  %We discuss the implications of our conclusions for the pursuit of theories of superuniversal quantum critical points. 
\end{abstract}
\bigskip

\newpage

\tableofcontents

\newpage

\section{Introduction}

In theories of non-relativistic particles in 2+1 dimensional flat spacetime, it is an established fact that attachment of even numbers of flux quanta to each particle does not change their statistics, provided the world lines of the particles do not intersect  \cite{Wilczek-1982}. 
This mapping from the original system of interacting particles to  an equivalent  system of  (also interacting) ``composite particles'' (fermions or bosons) 
coupled to a dynamical Abelian Chern-Simons gauge field is an identity at the level of their partition functions (see Ref. \cite{Fradkin-2013} 
for a review). % of these mappings). 
These mappings  have played a key role in the theory of the fractional quantum Hall fluids \cite{Jain-1989,Zhang-1989,Lopez-1991,Halperin-1993},  
in particular in elucidating their topological nature %of these fluids 
\cite{Wen-1990,Wen-1990b,Wen-1995,Moore-1991}, 
and showing that %the effective low energy theory of these topological fluids is a  Chern-Simons gauge theory \cite{Witten1989}.
they are described by a Chern-Simons gauge theory at low energies \cite{Witten1989}. 
With subtle but important differences, analogous mappings for relativistic quantum field theories %of massive particles 
in 2+1 dimensions 
between massive scalar fields and %massive
 Dirac fermions 
were argued by Polyakov \cite{Polyakov1988}. Because this duality involves transmutation of both statistics and spin, it does not accommodate the exact invariance under flux attachment seen in its non-relativistic counterpart.

Recently, a similar duality to Polyakov's was conjectured to hold, relating a Wilson-Fisher boson coupled to a Chern-Simons gauge field to one of a free Dirac fermion.
From this ``3D bosonization'' duality, it was shown that
one can derive a web of new dualities\footnote{See, also, the early approaches to duality in 2+1 dimensions in Refs. \cite{Burgess-Quevedo-1993,Fradkin1994,LeGuillou-1997}.} between relativistic quantum field theories in 2+1 
dimensions %\footnote{For the sake of concision, throughout this work we will refer to these Abelian dualities as ``the duality web.''} 
\cite{Seiberg2016,Karch2016}. 
These conjectures were motivated, in part, by the remarkable duality found between non-Abelian Chern-Simons gauge theories coupled to matter in the 't Hooft large-$N$ limit \cite{Aharony2016,Aharony2012,Giombi2012} and by Son's proposal to map the problem of the half-filled Landau level \cite{Halperin-1993} to a theory of massless Dirac fermions in 2+1 dimensions \cite{Son2015}, as well as the work connecting this problem to the theory of topological insulators in 3+1 dimensions %\cite{Senthil2013,Wang2014}
\cite{Wang2015,Metlitski2016}. The evidence for these dualities has been steadily mounting, with derivations from Euclidean lattice models \cite{Chen2017},  wire constructions \cite{Mross2016,Mross2017}, and deformations of supersymmetric dualities \cite{Kachru2015,Kachru2016a,Kachru2016}. %However, these dualities do not realize the invariance under flux attachment seen in their non-relativistic counterparts.
However, it has remained an open problem to construct derivations of these dualities in which relativistic flux attachment is implemented in a simple and transparent way explicitly using the Chern-Simons term. %In this context loop models, that will be introduced and discussed below, offer a natural way to approach this problem.

In this article, %we resolve this %tension between the point of view of the loop models of Eq. \eqref{eq: schematic loop model} and the local field theories \eqref{eq: schematic CFT} they seem to approach at criticality. 
%we answer this question. 
we show that such derivations can be constructed using relativistic models of current loops in 2+1 dimensions coupled to Chern-Simons gauge fields\footnote{Loop models have also been used by constructive field theorists to represent quantum field theories in 3+1 dimensional Euclidean spacetime  \cite{Symanzik1966,Symanzik1969,Frohlich-1982}.}. Such models can capture the physics of the theories of interest near criticality. They are analogues of models originally studied by Kivelson and one of us \cite{Fradkin1996}, which take the schematic form 
\be
\label{eq: schematic loop model}
\frac{1}{2}\left[g^2J^\mu\frac{1}{\sqrt{\pd^2}}J_\mu+2i\theta\epsilon^{\mu\nu\rho}J_\mu\frac{\pd_\nu}{\pd^2}J_\rho\right]\,,
\ee
where $J^\mu$ is a configuration of closed bosonic world lines satisfying $\pd_\mu J^\mu=0$. Here the first term is a long-ranged interaction of strength $g^2$, and the second term is a linking number which endows the matter with statistical angle $\theta$.   The model of Ref. \cite{Fradkin1996} displays self-duality under the modular group\footnote{PSL(2,$\mathbb{Z}$) is the group of all $2\times 2$ matrices with integer entries and unit determinant, defined up to an overall sign.} PSL$(2,\mathbb{Z})$ generated by particle-vortex duality,  which maps a theory of matter to one of vortices interacting with an emergent gauge field  \cite{Peskin1978, Dasgupta1981}, and flux attachment, which shifts $\theta$ by $\pi$. Similar PSL$(2,\mathbb{Z})$ structures arise in the study of the phase diagram of the quantum Hall effect \cite{Kivelson1992,Lutken1992,Lutken1993, Burgess2000, Hui2017} as well as in %other 
lattice models %in various dimensions which 
exhibiting oblique confinement \cite{thooft-1981,CARDY1982,CARDY1982a,Geraedts2012b,Geraedts2012c,GERAEDTS2013}. Another appears as electric-magnetic duality and $\Theta$-angle periodicity in 3+1 dimensions, which can be extended to correlation functions in 2+1 dimensional conformal field theories (CFTs) \cite{Witten2003,Leigh2003}. More recently, this modular group has appeared as a way of organizing the above mentioned web of 2+1 dimensional field theory dualities \cite{Seiberg2016}. %That PSL(2,$\mathbb{Z}$) has already 
%This structure has already 
%provided insight into the observed superuniversality--or sharing of critical exponents--of quantum Hall plateau transitions, leading to theories which appear to share correlation length exponents despite having distinct transport properties \cite{Hui2017}. 
It is important to note, however, that the PSL(2,$\mathbb{Z}$) of the duality web is not a group of dualities. Rather, it generates new dualities from known ones. On the other hand, the PSL(2,$\mathbb{Z}$) of the loop models we discuss here is to be taken as a group of dualities. 

The invariance under flux attachment appearing in Ref. \cite{Fradkin1996} is surprising given the apparent absence of such a symmetry in relativistic theories mentioned above. However, we will see that %the invariance of the loop models of Ref. \cite{Fradkin1996} under $\mathcal{T}$ 
this is a consequence of a choice of regularization which is impossible to apply to continuum Chern-Simons gauge theories coupled to matter. % with or without additional long-ranged interactions. 
This choice of regularization dispenses with the ``fractional spin" which massive particles are endowed with due to their interaction with the Chern-Simons gauge field. This fractional spin was at the center of Polyakov's original argument for boson-fermion duality, and it is responsible for the complete breaking of statistical periodicity\footnote{These issues with modular invariance do not arise in models consisting of two species of loops which can only statistically interact with one another. Such models display modular invariance and may be related to BF theories in the continuum limit \cite{Geraedts2012b,Geraedts2012c,GERAEDTS2013}.}. %Fractional spin is the avatar of the framing anomaly of these theories in a gapped phase, and it was at the center of Polyakov's early argument for boson-fermion duality \cite{Polyakov1988}. We then show that when fractional spin is included in the loop model, statistical periodicity is broken completely as a symmetry of the partition function.\footnote{These issues with modular invariance do not arise in models consisting of two species of loops which can only statistically interact with one another. Such models display modular invariance and may be related to BF theories in the continuum limit \cite{Geraedts2012b,Geraedts2012c,GERAEDTS2013}.} %While it does not affect local long-distance properties deep in a gapped phase (provided that the theory is on flat space), this breaking of periodicity will imply that the putative critical theories \eqref{eq: schematic CFT} are not dual to one another through statistical periodicity. %When the loop models are defined on the boundary of a 3+1 dimensional space, this will pave the way for an understanding of the action of PSL(2,$\mathbb{Z}$) on these models as the aforementioned PSL(2,$\mathbb{Z}$) of 3+1 dimensional electromagnetic duality and $\theta$-angle periodicity. 
%This breaking of periodicity in the loop models leads to behavior consistent with what we would expect for the CFTs of Eq. \eqref{eq: schematic CFT}, and we argue that it resolves the discrepancies in their universal properties. In other words, proper loop model descriptions of the theories defined by Eq. \eqref{eq: schematic CFT} must include fractional spin. We additionally provide an interpretation of fractional spin in the context of the duality web of Abelian Chern-Simons theories coupled to matter
Therefore, the inclusion of fractional spin allows contact with the Chern-Simons-matter theories comprising the web of dualities, enabling us to show that theories related by boson-fermion duality correspond to the same loop model. %: they have the same fractional spin factor. 
We are thus able to derive a duality web of loop models which parallels that of Refs. \cite{Seiberg2016,Karch2016}. %This can be thought of as an extension of Polyakov's argument to these dualities, and so is a simple, but nontrivial check on the duality web. 

We proceed as follows. In Section \ref{section: FK model}, we review the model of Ref. \cite{Fradkin1996}, discuss the inconsistency of statistical periodicity with the web of field theory dualities of Refs. \cite{Seiberg2016,Karch2016}, and review the appearance of PSL($2,\mathbb{Z}$) in both contexts. We then introduce the notion of fractional spin in Section \ref{section: fractional spin}, and we describe how it breaks statistical periodicity and is generic if our goal is to realize theories of relativistic matter coupled to Chern-Simons gauge fields at criticality. In Section \ref{section: duality web}, we show that the inclusion of fractional spin leads to consistency with the duality web and derive a parallel duality web of loop models. 
We conclude in Section \ref{section: discussion}.

%These proposals, combined with recent work on dualities in quantum field theories ....

%Nevertheless, the global phase diagram of the quantum Hall effect has the property that plateau transitions at different filling fractions are associated with the same \emph{superuniversal} critical exponents and are characterized by a relativistic dynamical exponent $z\approx 1$ \cite{Kivelson1992,Jain1990,Shimshoni1997} (\textbf{there are probably other things we should cite here, but I'll leave this to you to make sure that I don't miss something--HG}). %Despite this phenomenon, examples of critical quantum field theories which display this kind of superuniversality have been rare. 
%The theories governing these transitions are seemingly related by attachment of even numbers of flux quanta. Despite this, tractable superuniversal families of relativistic quantum field theories are few and far between. 

\section{Flux Attachment in a Self-Dual Loop Model}
\label{section: FK model}

\subsection{Model}

Motivated by 
the fact that all quantum Hall plateau transitions appear to have essentially the same critical exponents \cite{Kivelson1992,Jain1990,Shimshoni1997}, a phenomenon referred to as superuniversality,
Kivelson and one of us wrote down a model of current loops with long-ranged ($1/r^2$, where $r$ is the distance in 2+1 dimensional Euclidean spacetime) and statistical (linking number) interactions on a 3D Euclidean lattice displaying invariance under flux attachment ($\mathcal{T}$) and self-duality under particle-vortex duality ($\mathcal{S}$) \cite{Fradkin1996}. This model therefore describes superuniversal families of fixed points related by elements of the modular group generated by $\mathcal{S}$ and $\mathcal{T}$, PSL$(2,\mathbb{Z})$.
These fixed points have the surprising property that they not only share critical exponents, but also conductivities and other transport properties. 

The loop model of Ref. \cite{Fradkin1996} consists of integer-valued current loop variables $J_\mu$ representing the world lines of bosons on a 2+1 dimensional Euclidean cubic lattice,  with marginally long-ranged and statistical interactions. The partition function is 
\begin{equation}
Z=\sum_{\{J_\mu\}}\delta(\Delta_\mu J^\mu)e^{-S}\,,
\end{equation}
where the delta function enforces the condition that the currents $J_\mu$ are conserved, or that the world lines form closed loops. We require that the world lines are {\em non intersecting}, meaning that the bosons have a strong short-ranged repulsive interaction (``hard-core''). The action $S$ is defined to be
%\begin{widetext}
\begin{align}
S=&\frac{1}{2}\sum_{r,r'} J^\mu(r) G_{\mu\nu}(r-r')J^\nu(r')+\frac{i}{2}\sum_{r,R}J^\mu(r) K_{\mu\nu}(r,R)J^\nu(R)\nonumber\\
\label{eq: FK model}
&+i\sum_{r,r'}e(r-r')J^\mu(r)A_\mu(r')+\sum_{R,R'}h(R-R')\epsilon^{\mu\nu\rho}J_\mu(R)\Delta_\nu A_\rho(R')\\
&+\frac{1}{2}\sum_{r,r'}A_\mu(r)\Pi^{\mu\nu}(r,r')A_\nu(r')\,,\nonumber
\end{align}
%\end{widetext}
where $r,r'$ are sites on the direct lattice, $R$ are sites on the dual lattice, $\Delta_\mu$ is a right lattice derivative, and $A_\mu$ is a background probe electromagnetic field. Importantly, in this model we regard the loops as matter and flux world lines which follow each other, being separated by a rigid translation, so $R=r+(1/2,1/2,1/2)$. 

The symmetric tensor $G_{\mu\nu}$ and the antisymmetric tensor $K_{\mu\nu}$ are assumed to behave at  long distances such that in momentum space they take the form
\bea
G_{\mu\nu}(p)&=&\frac{g^2}{|p|}\left(\delta_{\mu\nu}-p_\mu p_\nu/p^2\right)\,, 
\label{eq:Gmunu}\\
K_{\mu\nu}(p)&=&2i\theta \; \epsilon_{\mu\nu\rho}\frac{p^\rho}{p^2}\,.
\label{eq:Kmunu}
\eea
Here $G_{\mu\nu}$ represents long-ranged interactions (i.e. a $1/r^2$ interaction, where $r$ is the Euclidean distance in 2+1 spacetime dimensions), and $K_{\mu\nu}$ represents the statistical interaction between matter fields (direct lattice) and flux (dual lattice) currents. In Eq. \eqref{eq:Gmunu}, the parameter $g$ is the coupling constant. The parameter $\theta$ of Eq. \eqref{eq:Kmunu} is the statistical angle of the world lines, and so $\theta/\pi$ is the number of flux quanta attached to each matter particle. 

%\begin{figure}
%\centering
%\includegraphics[width=0.1\textwidth]{self_linking_graphic.png}
%\caption{A self-linking process, in which a particle-antiparticle pair is created, braided, and subsequently annihilated.}
%\label{figure: self-linking}
%\end{figure}

Because we have defined matter and flux world lines to follow one another, conventional self-linking processes are absent. %of the type shown in Fig. \ref{figure: self-linking} are absent. % (by definition). 
As a result, the action for the statistical interaction of a given closed loop configuration $J_\mu$ is
\be
\theta\times \Phi[J],\quad \textrm{where}\; \Phi[J]\in2\mathbb{Z}\,,
\ee
where $\Phi[J]$ is \emph{twice} the linking number of the loop configuration. In the continuum limit, $\Phi[J]$ is given by
\bea
\Phi[J]&=&%-\int d^3x \epsilon^{\mu\nu\rho}J_\mu\frac{\pd_\nu}{\pd^2}J_\rho
\frac{1}{2\theta}\int \frac{d^3p}{(2\pi)^3}\; J_\mu(-p) \; K^{\mu\nu}(p)\; J_\nu(p)\\
&=&\frac{1}{4\pi}\int d^3x \int d^3y \; \epsilon^{\mu\nu\rho}\; J_\mu(x) \; \frac{(x_\nu-y_\nu)}{|x-y|^3} \; J_\rho(y)\nonumber
\eea
and is an even integer, so long as $J_\mu$ does not include any self-linking processes. This is twice the linking number since it counts each link twice (or each particle exchange once). The phase $\theta\Phi[J]$ should be regarded as the Berry phase of a configuration of closed loops labeled by the currents $J_\mu$.

%While seemingly awkward, this definition of the  Berry phase $\Phi$ will prove convenient in later sections when we discuss self-linking. 
In terms of $\Phi$, the partition function for long, closed loops can be written as
\be
Z=\sum_{\{J_\mu\}}\delta(\Delta_\mu J^\mu)\,e^{i\theta \Phi[J]}\, e^{-\frac{1}{2} \sum_{r, r'} J^\mu(r) G_{\mu\nu}(r-r')J^\nu(r')}\,.
\ee
where we have suppressed source terms. Because $\Phi[J]$ is an even integer, the partition function is invariant under %$2\theta\mapsto2\theta+2\pi$, or 
\be
\mathcal{T}:\theta\mapsto\theta+\pi\,.
\label{eq:Tau}
\ee
In other words, this theory is invariant under attachment of any number of flux quanta. Physically, this is due to our neglect of self-linking, which would correspond to single exchange processes, and so the exchange processes allowed in the theory come in pairs.\footnote{In Ref. \cite{Fradkin1996}, the statistical angle is defined (in the current notation) as $2\theta$. This was convenient in that work since self-linking processes were not allowed. Since we will relax this constraint soon, we have chosen to use the more conventional definition of the statistical angle as $\theta$.} Allowed exchange processes involving fermions therefore have the same amplitudes as their bosonic counterparts.

The reader may worry about the fact that we seem to allow $\theta$ to take fractional values. If we were to think of the statistical interaction as being obtained by integrating out a Chern-Simons gauge field, this would be inconsistent with gauge invariance in a purely 2+1 dimensional theory. Additionally, because $\theta\sim 1/k$, where $k$ is the Chern-Simons level, $\mathcal{T}$ transformations do not map integer levels to integer levels. This can be resolved by the introduction of auxiliary gauge fields so that no gauge field in the theory has a fractional level, as has been done in the study of the fractional quantum Hall effect (see e.g. Refs. \cite{Fradkin-1998,Fradkin-2013}). We will nevertheless proceed with fractional values of $\theta$ and $k$ for now since they should not affect local properties of the theory, and they do not run afoul of gauge invariance if the theory is defined on the boundary of a 3+1 dimensional system. %However, in Section \ref{section: duality web} the introduction of auxiliary gauge fields will prove useful. 
%We leave a detailed discussion of how we may properly define theories with arbitrary rational level to Appendix \ref{appendix: flux quantization}. %Since fractional levels are innocuous if we consider theories residing a 2+1-D surface in a 3+1-D bulk system, though, the reader may prefer to interpret the theory \eqref{eq: FK model} as living in such a setting. 

In addition to invariance  under shifts of the statistical angle of Eq. \eqref{eq:Tau}, it can easily be seen that the model in Eq. \eqref{eq: FK model} is also {\em self-dual} (in the absence of background fields, which break self-duality explicitly) under bosonic particle-vortex duality \cite{Peskin1978,Dasgupta1981}. This duality is a consequence of the fact that in 2+1 dimensions, a conserved current $J_\mu$ can be related to the field strength of an emergent gauge field $a_\mu$
\be
J_\mu=\frac{1}{2\pi}\epsilon_{\mu\nu\lambda}\Delta^\nu a^\lambda\,.
\ee
In the case of the 3D XY model, this allows one to rewrite the partition function as one of bosonic vortex variables strongly interacting via a logarithmic potential mediated by $a_\mu$. This dual theory is known as the Abelian Higgs model. In general, bosonic particle-vortex duality relates the symmetric, or insulating, phase of the matter variables to the broken symmetry, or superfluid, phase of the vortex variables: matter loops are scarce when vortex loops condense and vice versa.  For the models described by Eq. \eqref{eq: FK model}, particle-vortex duality is the map (see Appendix \ref{appendix: self-duality}),
\be
\mathcal{S}:\tau\mapsto-\frac{1}{\tau}\,,
\ee
where we have defined the modular parameter
\be
\tau=\frac{\theta}{\pi}+i\frac{g^2}{2\pi}\,.
\ee
Together, $\mathcal{S}$ and $\mathcal{T}$ generate the modular group PSL$(2,\mathbb{Z})$, which is the group of transformations
\be
\tau\mapsto\frac{a\tau+b}{c\tau+d}\,,
\ee
where $a,b,c,d\in\mathbb{Z}$ and $ad-bc=1$.

Invariance of the partition function under PSL$(2,\mathbb{Z})$ enabled the authors of Ref. \cite{Fradkin1996} to make  predictions for the DC conductivities of this theory at the so-called modular fixed point values of $\tau$. These are the points which are invariant under a particular modular transformation. We briefly review these results in the following subsection. In particular, since at the fixed points of PSL$(2,\mathbb{Z})$  the longitudinal conductivity $\sigma_{xx}$ is {\em finite}, the theory at such fixed points must also be at a fixed point in the sense of the renormalization group. %Which conformal field theory describes such fixed points is the main focus of the present paper.
One of the goals of the present work is to understand the nature of the conformal field theories describing these fixed points. 

\subsection{Modular Fixed Points and Superuniversal Transport}
\label{subsection: fixed points}

If we consider the partition function to be invariant under modular transformations in PSL$(2,\mathbb{Z})$, then we can fully constrain transport properties at the modular fixed points. Each fixed point can be related to one of $\tau=i$ (invariant under $\mathcal{S}$), $\tau=\frac{1}{2}+i\frac{\sqrt{3}}{2}$ (invariant under $\mathcal{T}\mathcal{S}$), $\tau=i\infty$, or $\tau=\infty$. %We will denote the set of points related to a modular fixed point $\tau_*$ by an element of PSL$(2,\mathbb{Z})$ with brackets as $[\tau_*]$. 

Under a modular transformation which leaves a fixed point invariant, we expect invariance of the loop-loop correlation function 
%\begin{widetext}
\be
D_{\mu\nu}(p;\tau)=\langle J_\mu(p)J_\nu(p)\rangle= D_{\mathrm{even}}(p;\tau)(\delta_{\mu\nu}-p_\mu p_\nu/p^2)+D_{\mathrm{odd}}(p;\tau)\epsilon_{\mu\nu\lambda}\frac{p^\lambda}{|p|}\,.
\ee
%\end{widetext}
Calculating $D_{\mu\nu}$ then amounts to writing down how it transforms under the modular transformation which leaves the fixed point invariant, equating that result to $D_{\mu\nu}$, and then solving. It is convenient to define 
\be
D(\tau)=\frac{2\pi}{|p|}(D_{\mathrm{odd}}(\tau)-iD_{\mathrm{even}}(\tau))\,.
\ee
One can derive the transformation law for $D(\tau)$ under $\mathcal{S}$ by exploiting the invariance of the %full 
current-current correlation function %with a particular background source $A_\mu$ 
\be
\mathcal{K}_{\mu\nu}=-\frac{\delta}{\delta A_\mu}\frac{\delta}{\delta A_\nu}\log Z[A]\big|_{A=0}\,,
\ee  
which is invariant under $\emph{any}$ duality transformation and tracks how the source terms transform. It is \emph{not} the same as the loop-loop correlation function, although they are related. Some algebra \cite{Fradkin1996} shows that the invariance of $\mathcal{K}_{\mu\nu}$ implies
\be
\label{eq: FK S on D}
D\left(-{1\over\tau}\right)=\tau^2D(\tau)+\tau\,.
\ee
This equation implies that $D(\tau)$ is not invariant under $\mathcal{S}$, instead transforming almost as a rank 2 modular form \cite{Fradkin1996}. We say almost because of the the last term in Eq. \eqref{eq: FK S on D}, which is known as the modular anomaly.

The conductivity,\footnote{In general, the conductivity is a function of the ratio of frequency and temperature \cite{Damle1997}. In this paper, we will exclusively consider optical conductivities, or the limit $T/\omega\rightarrow0$.} in units of $e^2/\hbar$, is defined in terms of the loop-loop correlation function as 
\begin{equation}
\sigma_{xx}(\tau)=\frac{1}{2\pi}\operatorname{Im}[D(\tau)]\,,\qquad  \sigma_{xy}(\tau)=\frac{1}{2\pi}\operatorname{Re}[D(\tau)]\,.
\end{equation}

This result enables us to immediately calculate the conductivity at the fixed point $\tau=i$, which is invariant under $\mathcal{S}$ transformations
\be
D(i)=-D(i)+i\Rightarrow D(i)=\frac{i}{2}\,,
\ee
so the conductivity at $\tau=i$ is
\be
\sigma_{xx}(i)=\frac{1}{2\pi}\operatorname{Im}[D(i)]=\frac{1}{4\pi},\qquad \sigma_{xy}(i)=0\,.
\ee
This gives a consistent result with continuum particle-vortex duality because it only requires self-duality under $\mathcal{S}$. The transport properties of this fixed point with Dirac fermion matter have been explored in detail in Ref. \cite{Hsiao2017}. %The results of that work are related to this bosonic result by the bosonization duality \cite{Mross2017}.

Before moving on to the other fixed points, a general result valid for all fixed points can be derived if we consider $D(\tau)$ to be invariant under $\mathcal{T}$ transformations % (this involves choosing a particular branch for $Z$ to live on in the complex plane parameterized by $g^2$ and $\theta$). 
\be
\label{eq: FK T on D}
D(\tau+1)=D(\tau)\,.
\ee
This can be thought of as a statement of superuniversality, as it equates conductivities at different values of $\theta=\pi\operatorname{Re}[\tau]$. It implies that the general transformation law for $D(\tau)$ is 
\be
D\left(\frac{a\tau+b}{c\tau+d}\right)=(c\tau+d)^2D(\tau)+c(c\tau+d)\,,
\ee     
%meaning that $D(\tau)$ is almost (because of the modular anomaly) a modular form of weight 2. 
This enables us to solve for $D(\tau)$ at an arbitrary fixed point. In particular, it enables us to uniquely determine $D(\tau)$ at the fixed points
\be
\label{eq: conductivity prediction from modular inv}
D(\tau)=\frac{i}{2\operatorname{Im}[\tau]}\,.
\ee
Notice that this implies that the Hall conductivity is fixed at zero. The only ingredient required to obtain this result is modular invariance, manifested in duality, Eq. \eqref{eq: FK S on D}, and periodicity, Eq. \eqref{eq: FK T on D}. In Ref. \cite{Fradkin1996}, this result was interpreted as implying that when the loop model is at a modular fixed point where $\sigma_{xx}$ is finite (e.g. $\tau=i$), it is at a critical point, where the loops become arbitrarily large and proliferate. In other words, at these modular fixed points the loop model must at a renormalization group  fixed point represented by a scale-invariant (and, presumably, conformally invariant) quantum field theory.\footnote{Up to subtle orders of limits, a 2+1-dimensional theory of charged fields at criticality, i.e. a CFT, can have a finite longitudinal conductivity in the thermodynamic limit. This finite dimensionless quantity is a universal property of the CFT.}

%We can use this to derive conductivities for the other fixed points. The point at $\tau=\frac{1}{2}+i\frac{\sqrt{3}}{2}$, referred to as the self-dual fermion point in Ref. \cite{Fradkin1996} despite having $\theta=\pi/2$, has conductivity $\sigma_{xx}(1/2+i\sqrt{3}/2)=\frac{1}{2\pi}\frac{\sqrt{3}}{3}$. Notice that the only ingredient required here to obtain this result is modular invariance, manifested in Eqs. \eqref{eq: FK S on D} and \eqref{eq: FK T on D}. In Ref. \cite{Fradkin1996}, this was interpreted as implying that the loop model is at a critical point at which the loops become arbitrarily large and proliferate. In other words, at these finite modular fixed points the loop model must at a renormalization group  fixed point represented by a scale-invariant (and, presumably, local and conformally invariant) quantum field theory.\footnote{Up to subtle orders of limits, a 2+1-dimensional theory of charged fields at criticality, i.e. a CFT, can have a finite longitudinal conductivity in the thermodynamic limit. This finite dimensionless quantity is a universal property of the CFT.}

We can use Eq. \eqref{eq: conductivity prediction from modular inv} to derive conductivities for the other fixed points. The point at $\tau=\frac{1}{2}+i\frac{\sqrt{3}}{2}$, referred to as the self-dual fermion point in Ref. \cite{Fradkin1996} despite having $\theta=\pi/2$, has conductivity $\sigma_{xx}(1/2+i\sqrt{3}/2)=\frac{1}{2\pi}\frac{\sqrt{3}}{3}$. Additionally, in Ref. \cite{Fradkin1996}, it was noted that there are modular fixed points on the real axis, which formally have $\sigma_{xx}(\infty)\rightarrow\infty$. However, in this limit, where the parity-even long-ranged interactions vanish, the short-ranged interactions can no longer be neglected  and, in a sense, become dominant. In the next subsection, we will see that these ``pathological'' fixed points of the modular symmetry are in conflict with results derived from the duality web. We will later see in Section \ref{section: fractional spin} that the correct definition of the loop models at short distances necessarily implies fractional spin, which spoils the periodicity symmetry (and hence modular invariance).

%Similarly, the real axis fixed points have %$\operatorname{Im}[\tau]=0$. If these loop models could be described by Eq. \eqref{eq: continuum CFT}, then they would correspond to Chern-Simons-matter theories without long-ranged interactions. Eq. \eqref{eq: conductivity prediction from modular inv} would then tell us that these theories have
%$\sigma_{xx}(\infty)\rightarrow\infty$. 

%a consequence of the fact that, in the absence of long-ranged interactions, current loops proliferate unsuppressed. Short-ranged interactions might then be required to make the conductivity finite and ensure convergence of the partition function as criticality is approached, and the partition sums over loops of arbitrarily large sizes. 
%However, the only ingredient required here to obtain this result is modular invariance, manifested in Eqs. \eqref{eq: FK S on D} and \eqref{eq: FK T on D}. %We therefore might expect that short-ranged interactions spoil modular invariance in some way. In Section \ref{section: fractional spin}, we will see that this is indeed the case. %First, however, we will show that this result for the conductivity directly conflicts with the web of 2+1-D CFT dualities of Refs. \cite{Seiberg2016,Karch2016}. %and that, in fact, the physics which breaks modular invariance is even more generic than it would be if it were exclusively the consequence of dangerously irrelevant short-ranged interactions

\subsection{Modular Invariance and the Web of Dualities}
\label{subsection: modular transformations and the web}

\subsubsection{An Attempt at a Field Theory Description}

It is natural to ask whether at criticality the loop models in Eq. \eqref{eq: FK model} approach relativistically invariant CFTs which inherit modular invariance and what the interpretation of this might be in the context of the duality web of Refs. \cite{Seiberg2016,Karch2016} and its own PSL($2,\mathbb{Z})$ structure. These theories would display superuniversality in both critical exponents \emph{and transport}. The only obvious local candidates for such theories would consist of matter fields on a 2+1 dimensional surface in a bulk 3+1 dimensional spacetime interacting via an emergent, dynamical gauge field that propagates in the bulk with Maxwell and $\Theta$ terms. Such theories are  analogous to models of fractional topological insulators \cite{Zhang2010,Zhang2012,Maciejko2015,Swingle2011,Chan-2013,Ye2014,Ye2016,Cho2017,Ye2017}, which in the bulk have fractional $\Theta$-angles and support gapless matter on their boundaries. The connection to the loop model Eq. \eqref{eq: FK model} is immediate: the Maxwell term would then integrate to the surface as long-ranged $1/r^2$ interactions between the matter particles, and the $\Theta$ term would become a Chern-Simons term of level $k=\Theta/2\pi$ which endows the matter with fractional statistics. One may also consider the surface theory on its own without a bulk, but this theory would be non-local. Without referencing a bulk, the Lagrangian for these theories takes the form
\be
\label{eq: schematic CFT}
\mathcal{L}_{\mathrm{CFT}}=\mathcal{L}_{\mathrm{matter}}[a]-\frac{1}{4e^2}f^{\mu\nu}\frac{i}{\sqrt{\pd^2}}f_{\mu\nu}+\frac{k}{4\pi}ada\,,
\ee
where $a_\mu$ is a dynamical gauge field, $f_{\mu\nu}=\pd_\mu a_\nu-\pd_\nu a_\mu$, we use the notation $AdB=\epsilon^{\mu\nu\rho}A_\mu\pd_\nu B_\rho$, and we have again suppressed background terms. $\mathcal{L}_{\mathrm{matter}}[a]$ can be taken to be the Lagrangian either for a single species of Dirac fermion or  Wilson-Fisher boson coupled to $a_\mu$. In the case of bosonic matter, a natural modular parameter for this theory is $k+i\frac{2\pi}{e^2}=\frac{\Theta}{2\pi}+i\frac{2\pi}{e^2}$, which would correspond to $-\frac{1}{\tau}$ in the loop model language.

Models of the form Eq. \eqref{eq: schematic CFT} are self-dual under $\mathcal{S}$, which can be taken to be (fermionic or bosonic) particle-vortex duality \cite{Peskin1978,Dasgupta1981,Metlitski2016,Wang2015}. Recently,  inspired by the web of field theory dualities, this self-duality has been explored anew \cite{Mross2017,Hsiao2017}, building on the earlier analytic work on bosonic loop models in Ref. \cite{Fradkin1996} and on numerical work at $\theta=0$ \cite{Geraedts2012}. However, invariance under $\mathcal{T}$ is far from manifest in these theories. It is a deformation of the Chern-Simons level, which does not preserve the phase diagram of the theory, affecting both local (e.g. Hall conductivities) and global (e.g. ground state degeneracy on a torus) properties of the gapped phases. Moreover, invariance of transport properties under $\mathcal{T}$ leads to predictions which are inconsistent with those of the duality web, which does not accommodate sharing of transport properties amongst theories with general values of $k$, as we will see below in the next subsection. Theories of the form of Eq. \eqref{eq: schematic CFT} related by $\mathcal{T}$ therefore cannot be dual. In Section \ref{section: fractional spin}, we will see that this apparent tension is resolved upon the introduction of fractional spin, which breaks periodicity in the loop models completely. %, although it may be the case that they share certain universal properties. This inspires the question: what the relationship is between the conformal field theories of Eq. \eqref{eq: schematic CFT} and the loop models of Eq. \eqref{eq: schematic loop model}?
%Clearly, then, we should not expect $\mathcal{T}$-transformations to leave the partition functions of Chern-Simons-matter theories invariant. 

\subsubsection{Inconsistency of Modular Invariance with the Duality Web}
\label{subsubsection: conflict with duality web}

We can check for consistency of the transport predictions one obtains from modular invariance with those from the duality web of Refs. \cite{Seiberg2016,Karch2016}. While the predictions for the modular fixed point at $\tau=i$ are consistent whether the matter content is fermionic or bosonic \cite{Geraedts2012,Hsiao2017}, this is not the case for the fixed points on the real $\tau$ line. We can see this by studying the (conjectured) duality between a free Dirac fermion\footnote{In this paper, we approximate the $\eta$-invariant by $\frac{1}{8\pi}AdA$ and include it in the action.} and a gauged Wilson-Fisher fixed point
%\begin{widetext}
\be
\label{eq: free fermion duality}
i\bar{\psi}\slashed{D}_A\psi-\frac{1}{8\pi}AdA\longleftrightarrow|D_a\phi|^2-|\phi|^4+\frac{1}{4\pi}ada+\frac{1}{2\pi}adA\longleftrightarrow|D_{b-A}\tilde{\phi}|^2-|\tilde{\phi}|^4-\frac{1}{4\pi}bdb\,,
\ee
%\end{widetext}
where $A$ is a background gauge field, $D_{\alpha}^\mu=\pd^\mu-i\alpha^\mu$, and we use the notation $\slashed{\alpha}=\alpha^\mu\gamma_\mu$, where the $\gamma_\mu$'s are the Dirac gamma matrices. Throughout this work, we will use $\longleftrightarrow$ to indicate duality. The duality between the bosonic theories is a particle-vortex duality. We expect the bosonic theories to correspond to the loop model fixed points at $\tau=\mp1$, $k=\pm1$ (invariant under $\mathcal{T}^2\mathcal{S}$), where $k$ is the level of the Chern-Simons gauge field. 

Because the Dirac fermion on the left hand side of Eq. \eqref{eq: free fermion duality}  is free, we can use this duality to calculate the optical conductivities of the strongly coupled bosonic theories. As with the loop models, duality implies that the  correlation function,
\be
\mathcal{K}_{\mu\nu}=-\frac{\delta}{\delta A_\mu}\frac{\delta}{\delta A_\nu}\log Z[A]\big|_{A=0}\,,
\ee
should be the same for each of these theories. From the free fermion theory, it is easy to calculate the conductivity (again in units of $e^2/\hbar$)
\be
\frac{1}{i\omega}\mathcal{K}_{xx}=\frac{1}{16},\,\frac{1}{i\omega}\mathcal{K}_{xy}=-\frac{1}{4\pi}\,.
\ee
From the particle-vortex duality in Eq. \eqref{eq: free fermion duality}, we see that the current-current correlation functions for the $\phi$ and $\tilde{\phi}$ gauge currents, $J_\mu$ and $\tilde{J}_\mu$ respectively, differ only in the Hall conductivity
\begin{align}
\frac{1}{i\omega}\mathcal{K}_{ij}(\omega)&=\frac{1}{i\omega}\langle\tilde{J}_i(-\omega)\tilde{J}_j(\omega)\rangle\nonumber\\
&=\frac{1}{i\omega}\langle J_i(-\omega)J_j(\omega)\rangle-\frac{1}{2\pi}\epsilon_{ij}.
\end{align}
Notice that this matches Eq. \eqref{eq: FK S on D} for the case $\tau=-1$. Upon denoting\footnote{Usually, in gauge theories one is interested in the 1PI conductivity, which is defined using the polarization tensor $\Pi_{ij}(\omega)$ as $\sigma^{\mathrm{1PI}}_{ij}=\Pi_{ij}(\omega)/i\omega$ and characterizes the current response to the sum of the probe and emergent electric fields. In general, these conductivities map to resistivities under particle-vortex duality \cite{Geraedts2012b,Geraedts2012c,Herzog2007,Hsiao2017,Goldman2017}. However, for correct comparison to the constraints obtained in the loop model of Ref. \cite{Fradkin1996}, we use the conductivities associated with the full current-current correlation functions.} 
\begin{align}
 \sigma^\phi_{ij}&=\frac{1}{i\omega} \langle J_i(-\omega)J_j(\omega)\rangle\,,\\
 \sigma^{\tilde\phi}_{ij}&=\frac{1}{i\omega} \langle\tilde{J}_i(-\omega)\tilde{J}_j(\omega)\rangle\,,
 \end{align}
we obtain 
\be
\sigma_{xx}^\phi=\sigma_{xx}^{\tilde{\phi}}=\frac{1}{16}\,,\qquad \sigma_{xy}^\phi=-\sigma_{xy}^{\tilde{\phi}}=\frac{1}{4\pi}\,.
\ee
This result disagrees with the prediction of Eq. \eqref{eq: conductivity prediction from modular inv} of modular invariance! Not only is the Hall conductivity nonvanishing, but the transverse conductivity is finite. Hence, if the bosonization duality is to be trusted, periodicity cannot extend to transport in theories of gapless matter coupled to a Chern-Simons gauge field, i.e. a version of Eq. \eqref{eq: FK T on D} cannot hold. %Since the conductivity predictions for the fixed point which is self-dual under $\mathcal{S}$ ($\tau=i$) agree between the CFT dualities and the loop model, 
The conclusion one is driven toward is that any loop model description of these theories cannot be periodic either.

\subsubsection{A Multitude of Modular Groups}
Before concluding this section, we note the appearance of PSL($2,\mathbb{Z}$) in the context of the duality web of Ref. \cite{Seiberg2016} in order to distinguish it from the modular group we are primarily concerned with. In that work, new dualities are obtained from old ones by the application of modular transformations to the conformal field theories on either side of a duality. If $\Phi$ denotes a set of dynamical fields and $A$ is a background gauge field, these transformations act on a Lagrangian $\mathcal{L}[\Phi,A]$ as \cite{Witten2003}
\bea
\tilde{\mathcal{S}}&:&\mathcal{L}[\Phi,A]\mapsto\mathcal{L}[\Phi,a]+\frac{1}{2\pi}Ada\,,\\
\tilde{\mathcal{T}}&:&\mathcal{L}[\Phi,A]\mapsto\mathcal{L}[\Phi,A]+\frac{1}{4\pi}AdA\,,
\eea
where $a$ is a dynamical gauge field. Here $\tilde{\mathcal{S}}$ involves gauging $A\rightarrow a$ and adding a BF term coupling $a$ to a new background gauge field (also denoted $A$), and $\tilde{\mathcal{T}}$ is simply the addition of a background Chern-Simons term. If $A$ is allowed to exist in a bulk 3+1 dimensional spacetime for which $\mathcal{L}[\Phi,A]$ is the boundary Lagrangian, $\tilde{\mathcal{S}}$ and $\tilde{\mathcal{T}}$ correspond respectively to electromagnetic duality and $\Theta$-angle periodicity of the bulk theory. The modular group generated by $\tilde{\mathcal{S}}$ and $\tilde{\mathcal{T}}$ also organizes the global phase diagram of the fractional quantum Hall effect, where it has a natural action on the conductivities of the incompressible phases \cite{Kivelson1992,Lutken1992,Burgess2000,Lutken1993}. It has also provided insight into the problem of superuniversality, relating theories which appear to share correlation length exponents despite having distinct transport properties \cite{Hui2017}. 

In general, the modular transformations $\tilde{\mathcal{S}}$ and $\tilde{\mathcal{T}}$ \emph{are not} duality transformations themselves: they do not always leave the partition function of a particular theory invariant. This is obvious for $\tilde{\mathcal{T}}$, which shifts the Hall conductivity\footnote{Despite the fact that it is not a duality transformation, $\tilde{\mathcal{T}}$ does not change the fractional part of the Hall conductivity, which an universal obsevable in a topological phase \cite{Kivelson1992,Wen-1995} and in a CFT \cite{Closset2012}. This statement can be thought of as the 2+1 dimensional analogue of $\Theta$-angle periodicity in 3+1 dimensions, although it is important to emphasize that this is a statement about Chern-Simons terms of \emph{background} gauge fields.}. $\tilde{\mathcal{S}}$, on the other hand, is only occasionally a duality transformation, e.g. in the case of the duality between the Abelian Higgs model and a boson at its Wilson-Fisher fixed point. 
In contrast, the PSL($2,\mathbb{Z}$) associated with the self-dual loop model of Ref. \cite{Fradkin1996} is a group of dualities: there $\mathcal{S}$ is identified with particle-vortex duality, and $\mathcal{T}$ is periodicity. $\mathcal{S}$ can also be related to bulk electromagnetic duality in the case where $A$ is a dynamical field, albeit in a way slightly different from $\tilde{\mathcal{S}}$ \cite{Hsiao2017}.

\section{Fractional Spin and the Fate of %Statistical 
Periodicity}
\label{section: fractional spin}
%\subsection{Models}

%\subsection{Self-Linking, Fractional Spin, and the Fate of Modular Invariance}
\subsection{Fractional Spin and the Framing Anomaly}

\begin{comment}
A very natural question at this point is whether there is a continuum conformal field theory (CFT) which has the same modular self-duality for which a version of the above lattice model is a good UV completion. The $\mathcal{S}$-transformations, particle-vortex duality, indeed carries over to continuum theories of Wilson-Fisher scalar (or free Dirac fermion) matter interacting with a gauge field with the nonlocal propagator $G_{\mu\nu}(p)+iK_{\mu\nu}(p)$ \cite{Mross2017,Hsiao2017},
\be
\label{eq: continuum CFT}
S=\int d^3x \left(|D_a\phi|^2-|\phi|^4-\frac{1}{4g^2}f^{\mu\nu}\frac{i}{\sqrt{\pd^2}}f_{\mu\nu}+\frac{1}{4\theta}ada\right)
\ee
Such theories can be made local if the gauge field is allowed to propagate in a 3+1D bulk spacetime. However, whether a version of $\mathcal{T}$ invariance exists for such theories is far from obvious. In fact, it has been argued in the past that such invariance cannot appear at all in a local CFT \cite{Herzog2007}, although such arguments have not been ironclad. At the very least, any continuum CFTs we study should contain the aforementioned self-linking processes, so we can hope for at most $\mathcal{T}^2$ invariance, which shifts the statistical angle $\theta\mapsto\theta+2\pi$.
\end{comment}

The full PSL$(2,\mathbb{Z})$ invariance of the loop model of Ref. \cite{Fradkin1996} above relies on the absence of self-linking and thus of fractional spin. This corresponds to a convenient choice of regularization, but we will find that such regularization is not available for CFTs of the form of Eq. \eqref{eq: schematic CFT}. To see this, we must carefully include self-linking in the loop models reviewed above, as such processes generically appear in continuum field theories. Moreover, whenever one considers self-linking processes in a Chern-Simons theory, they are confronted with the framing anomaly, with which fractional spin is associated. We will find that (1) the inclusion of self-linking processes while neglecting fractional spin breaks the $\mathcal{T}$-invariance of the loop model not only down to invariance under $\mathcal{T}^2$ (the usual statistical periodicity $\theta\sim\theta+2\pi$), and that (2) fractional spin breaks $\mathcal{T}$-invariance entirely. 

Consider the loop model of Eq. \eqref{eq: FK model} with the inclusion of self-linking processes. For convenience, now and in the remainder of this article we will use a continuum description, replacing lattice sums with integrals. The reader may be concerned that this passage to the continuum is too cavalier. However, starting from a continuum, gapped field theory, we can always rewrite the partition function as a world line path integral without referencing a lattice. See, for example, Refs. \cite{Fradkin-2013,Polyakov:1987ez,PolyakovLesHouches1988,Grundberg1990,SHAJI1990}. %Indeed, we will gain a great deal of mileage from this fact in the second half of this work.

The linking number term in the action is
\begin{align}
\label{eq: linking number}
\theta\, \Phi[J]&=\theta\int d^3xJd^{-1}J=\frac{\theta}{4\pi}\int d^3x \int d^3y\,\epsilon^{\mu\nu\rho}J_\mu(x)\frac{(x_\nu-y_\nu)}{|x-y|^3}J_\rho(y)\,.
\end{align}
Note that to properly define this term, we must assume that the configuration $J$ does not involve any loops which cross. This constraint can be implemented through additional short-ranged interactions like those which characterize the Wilson-Fisher fixed point. Now consider a configuration of two loops, $J(x)=\ell^{(1)}(x)+\ell^{(2)}(x)$, where each $\ell^{(i)}$ represents a single closed loop with unit charge. The action of this configuration is 
\be
\label{eq: general linking number action}
\theta\,\Phi[J]=\theta\left(2\varphi[\ell^{(1)},\ell^{(2)}]+\varphi[\ell^{(1)},\ell^{(1)}]+\varphi[\ell^{(2)},\ell^{(2)}]\right)\,,
\ee
where
\be
\varphi[\ell^{(i)},\ell^{(j)}]=\frac{1}{4\pi}\int d^3xd^3y\,\epsilon^{\mu\nu\rho}\ell^{(i)}_\mu(x)\frac{(x_\nu-y_\nu)}{|x-y|^3}\ell^{(j)}_\rho(y)\,.
\ee
The first term in Eq. \eqref{eq: general linking number action} is twice the linking number of the two loops and is an integer-valued topological invariant: it simply counts the number of times the two loops link. This is the only term which appears in the model discussed in Section \ref{section: FK model}. The last two terms are referred to as the {\em writhes} of $\ell^{(1)}$ and $\ell^{(2)}$ respectively, denoted below as $W[\ell^{(i)}]=\varphi[\ell^{(i)},\ell^{(i)}]$. 

The writhe contains a ``fractional spin'' term, which Polyakov showed can transmute massive scalar bosons to massive Dirac fermions in 2+1-dimensions \cite{Polyakov1988}. Unlike the linking number, the writhe is not a topological invariant: it depends on the metric. It also generically breaks invariance under orientation-reversal of the loops, which can be thought of as particle-hole (or charge conjugation) symmetry ($\mathbf{PH}$), in addition to time-reversal ($\mathbf{T}$) and parity\footnote{Reflection about one of the spatial axes.} ($\mathbf{P}$). This metric dependence can in principle be eliminated by calculating self-linking numbers using a point-splitting regularization following Witten \cite{Witten1989}, in which the loops are broadened into ribbons with a framing vector $a_{\mathrm{f}}\hat{n}$, the edges of which having a well defined linking number $SL$,
\begin{align}
\label{eq: SL}
SL[\ell]=&\lim_{a_{\mathrm{f}}\rightarrow0}\frac{1}{4\pi}\oint_\ell dx^\mu\oint_\ell dy^\nu\epsilon_{\mu\nu\rho}\frac{(x^\rho-y^\rho+a_{\mathrm{f}}\hat{n}^\rho)}{|x-y+a_{\mathrm{f}}\hat{n}|^3}\,,
\end{align}
However, this is at the cost of introducing a framing ambiguity in the calculation of this linking number: there is in general no canonical way to convert a loop into a ribbon. On the other hand, we can break the topological character of the theory along with $\mathbf{PH}$, $\mathbf{T}$, and $\mathbf{P}$ by including the fractional spin, eliminating the framing ambiguity. This choice is the manifestation of the framing anomaly \cite{Witten1989,bar-natan1991} in the language of loop models.

\subsubsection{Self-Linking Without Fractional Spin: Point Splitting}
Let us consider what happens if we choose Witten's point-splitting procedure, which looks appealing because we may replace the writhe with a topological invariant. If we replace the writhe with the self-linking number and plug this into the action for the linking of two loops in Eq. \eqref{eq: general linking number action}, we obtain the action,
\be
S=\theta(SL[\ell^{(1)}]+SL[\ell^{(2)}])+2\theta\varphi[\ell^{(1)},\ell^{(2)}]\,.
\ee
The self-linking number $SL$ can take any integral value, so here $S$ is only invariant mod $2\pi$ under 
\be
\mathcal{T}^2:\theta\mapsto\theta+2\pi\,.
\ee
Here $e^{i\theta}$ is the phase the wave function picks up upon a single exchange process of two particles, as discussed in the previous section. It is worth noting that exchange processes which form closed loops %as in Fig. \ref{figure: self-linking} 
are only possible in relativistic theories, where we have particles and antiparticles available for braiding. In non-relativistic systems, to obtain this type of process, %something of the form of Fig. \ref{figure: self-linking}, 
one must compactify time and wrap the particle world lines around the time direction. %e.g. a process in which particle-antiparticle pair is created, exchanged, and subsequently annihilated, so $\theta$ is the statistical angle we are all familiar with. 

We have now found that the $\mathcal{T}$ invariance of a model without self-linking is broken down to the usual periodicity of the statistical angle when self-linking, but not fractional spin, is included. This means that the PSL$(2,\mathbb{Z})$ modular invariance of the model of Ref. \cite{Fradkin1996} is broken down to a %the subgroup $\Gamma_{\mathcal{S}}(2)$, which is 
subgroup generated by particle-vortex duality ($\mathcal{S}$) and $\mathcal{T}^2$. We may therefore be inclined to accept the framing ambiguity and proceed by calculating the partition function with a point-splitting regularization of the linking integral. However, we will soon see that not even this symmetry can be accommodated by the continuum CFTs we might hope to describe. 

\subsubsection{Introducing Fractional Spin}

Now consider the regularization in which the full writhe remains in the action without adopting a point-splitting regularization, following Polyakov. In this case, the action is frame independent, but this comes at the cost of reintroducing the metric. There is a general relation in knot theory relating $W[\ell]$ and $SL[\ell]$ \cite{White1969},
\be
W[\ell]=SL[\ell]-T[\ell]\,,
\ee
where $T[\ell]$ is referred to as the twist of the world line $\ell$. It is Polyakov's fractional spin term, and it can be written as \cite{Polyakov1988}
\be
\label{eq: twist}
T[\ell]=\frac{1}{2\pi}\oint_\ell\, \omega'=\frac{1}{2\pi}\oint_{\ell}ds\,\hat{e}\cdot(\hat{n}\times\pd_s\hat{n})\,,
\ee  
where $\ell$ is parameterized by the variable $s\in[0,L]$, $\hat{e}$ is the unit tangent vector to $\ell$, and $\hat{n}$ is again a chosen frame vector normal to $\ell$. This integral clearly depends on the metric, and it measures the angular rotation of $\hat{n}$ about $\hat{e}$. $\omega'$ is  the angular velocity of $\hat{n}$. It can be thought of as a spin connection restricted to $\ell$. However, this integral need not vanish on a flat manifold, and it can take non-integer values because it depends on the embedding of $\ell$ in spacetime.\footnote{For an explicit example, see Ref. \cite{Grundberg1990}}

Up to addition by an integer, the integral of Eq. \eqref{eq: twist} can be written as a Berry phase by extending $\hat{e}$ to a disk: $\hat{e}(s)\rightarrow\hat{e}(s,u)$, where $u\in[0,1]$ and $\hat{e}(s,u=1)=\hat{e}(s),\hat{e}(s,u=0)=\hat{e}_0\equiv\mathrm{constant}$,
\be
\label{eq: twist as Berry phase}
T[\ell]=\frac{1}{2\pi}\int_0^L ds\int_0^1du\,\hat{e}\cdot(\pd_s\hat{e}\times\pd_u\hat{e})+n,\,n\in\mathbb{Z}\,.
\ee
This Berry phase form is what earns this term the name of fractional spin. 

For $\theta=\pm\pi$, Polyakov argued that the loop model partition function for particles with this Berry phase (massive charged scalar bosons coupled to a Chern-Simons gauge field) is that of a single massive Dirac fermion of mass $M$, \footnote{Here, and in several places below, we do not make explicit the fact that loops have strong short range repulsive interactions, without which these expressions involving linking numbers  do not make sense.}
\begin{align}
\label{eq: Polyakov duality without gauge fields}
Z_{\mathrm{fermion}}&=\det[i\slashed{\pd} - M]=\int\mathcal{D}J\,\delta(\pd_\mu J^\mu)\, e^{-|m|L[J]- i\sgn(M)\pi\Phi[J]}\,.
\end{align}
%where $L[J]$ is the sum of the lengths of the loops in the configuration $J$ and $\Phi[J]$ is the linking number of Eq. \eqref{eq: linking number} of $J$ (without appeal to point splitting), which contains the full writhes and, therefore, the fractional spin factor of each loop. Its sign is determined by the sign of the fermion mass in the phase in question. Note that $J$ is not coupled to a gauge field here: this leads to the appearance of the parity anomaly, which breaks $\mathbf{P}, \mathbf{T}$, and $\mathbf{PH}$ even in the approach to criticality $m\rightarrow0$. We will discuss how to couple $J$ to a gauge field in detail in Section \ref{section: duality web}. 
This relation\footnote{An unpublished work by Ferreir{\'os} and one of us \cite{Ferreiros-2018} discusses an extension of Polyakov's duality in curved spacetimes.} requires some unpacking, especially since we will encounter several more like it in Section \ref{section: duality web}. Here we have fully passed to a continuum picture where $J^\mu$ is a current density and so is not restricted to be an integer (although $\int_{S}d\Sigma_\mu J^\mu\in\mathbb{Z}$ for any closed surface $S$), thus the use of $\int\mathcal{D}J$ rather than $\sum_{\{J\}}$. $L[J]$ is the sum of the lengths of the loops in the configuration $J$, and the term $-|m|L[J]$ represents tuning away from criticality into a phase of small loops so that the partition function converges. It is generic in loop models, despite the fact that we have suppressed it thus far. $\Phi[J]$ is the linking number \eqref{eq: linking number} of $J$ (without appeal to point splitting), which contains the full writhes and, therefore, the fractional spin factor of each loop. The sign of the linking number term matches the sign of the fermion mass $M$, which is proportional to $m$. Note that $J$ is not coupled to a gauge field here: this leads to the appearance of the parity anomaly, which breaks $\mathbf{P}, \mathbf{T}$, and $\mathbf{PH}$ even in the approach to criticality $m\rightarrow0$. We will discuss how to couple $J$ to a gauge field in detail in Section \ref{section: duality web}. For a review of Polyakov's original argument, see Appendix \ref{appendix: Polyakov}. See also later work fleshing out some of the details, Refs. \cite{Alekseev1988, Grundberg1989,Grundberg1990,SHAJI1990,Orland1988}.

For general statistical angle $\theta$, we have the loop model partition function 
\be
Z=\int\mathcal{D}J\,\delta(\pd_\mu J^\mu)\,e^{-|m|L[J]+i\theta\Phi[J]}\,.
\ee
Unless $\theta=0,\pm\pi$, $\mathbf{T}$, $\mathbf{P}$, and $\mathbf{PH}$ are broken explicitly even in the $m\rightarrow0$ limit. Since the value of the twist $T[\ell]$ is not restricted to the integers, $\Phi[J]$ is not restricted to the integers either. This means that fractional spin eliminates even the $\mathcal{T}^2$ symmetry we found via point-splitting. This has consequences for universal physics: for example, in the case $\theta=\pi$, the theory of a free Dirac fermion one obtains with fractional spin has a different correlation length exponent from the theory of spinless fermions one would have obtained neglecting fractional spin.

If periodicity is broken in the presence of fractional spin, how should one interpret shifts of the statistical angle? Say that we start with $\theta=\pi$, or the free Dirac fermion. When $\theta$ is shifted, the fractional spin is in turn shifted, and we can no longer make the mapping to a free Dirac fermion. It cannot be a theory of a higher spin particle either, since there are no non-trivial higher spin particles in 2+1-dimensions.\footnote{The reader can convince themselves of this by writing down the action for a Rarita-Schwinger (spin-$3/2$) field in 2+1 dimensions and considering its equations of motion. They will find that such a field does not propagate.} However, in Section \ref{section: duality web}, we will argue that the theory which can reproduce the same spin factor is a theory of Dirac fermions strongly coupled to Chern-Simons gauge fields. 

%There is a subtle issue that it is important to stress at this point. At the beginning of this Section, we assumed that the loops do not cross and that this can be implemented through short-ranged interactions. If we take the short-ranged interactions to arise from scalar self-interactions associated with a Wilson-Fisher fixed point, then we might take Polyakov's bosonization argument to suggest the gapless duality between a gauged Wilson-Fisher fixed point and a free Dirac fermion discussed in Section \ref{subsection: modular transformations and the web}. 
%However, there has thus far been no serious analysis of the role of short-ranged interactions in deriving bosonization dualities of loop models. 
%Moreover, another duality between a Gross-Neveu fixed point and a gauged free scalar fixed point is also thought to hold \cite{Aharony2012,Giombi2012,Karch2016}, but it is not clear to us how to obtain this duality from the perspective of loop models. 
%These difficulties with Polyakov's argument are important because the derivation of 3D bosonization by studying loop models is essentially a ultraviolet (UV) construction, as one is interested in the transmutation of gapped particles, so one really should not neglect short-ranged interactions. We leave this for future work.

\subsection{Fractional Spin is Generic}

Having established that the introduction of fractional spin breaks periodicity of the statistical phase completely, we now describe how fractional spin is a generic feature of loop models of the form of Eq. \eqref{eq: FK model}. %This will lead us to conclude that CFTs of the form \eqref{eq: schematic CFT} do not admit a regularization which dispenses with fractional spin. 
It is known that, in the presence of ultraviolet (UV) scales, Witten's point-splitting regularization described above does not generally eliminate fractional spin. If we decouple the $J$ variables by introducing an emergent gauge field $a$, this can be seen if we turn on arbitrarily weak short-ranged interactions in the form of a Maxwell term\footnote{The same thing happens if we UV complete the Chern-Simons terms to lattice fermions coupled to dynamical gauge fields, as was done in Ref. \cite{Chen2017} to construct a lattice proof of the duality between a gauged Wilson-Fisher boson and a free Dirac fermion.} 
\be
\mathcal{L}_{\mathrm{Maxwell}}=-\frac{1}{4g_\mathrm{M}^2}f^2\,.
\ee
Due to the existence of this term point-splitting no longer has the desired effect: one continues to obtain the metric-dependent $W[\ell]$ rather than the topological invariant $SL[\ell]$ \cite{HANSSON198992}. This is because the Maxwell term introduces a short-distance cutoff $a_{\mathrm{M}}=2\pi(g_\mathrm{M}^2k)^{-1}$, where $k$ is again the level of the Chern-Simons term, and the different result obtained by point-splitting is a consequence of the short-distance singularity of the Chern-Simons propagator in the absence of a natural cutoff. More physically, with the Maxwell term, flux is no longer localized on the matter world lines, but is smeared around the world line out to lengths of order $a_{\mathrm{M}}$. When this singularity is smoothed out, the self-linking number becomes metric-dependent but frame independent, leading to the full writhe. The existence of a Maxwell term therefore removes the UV ambiguities that exist in pure Chern-Simons theory and renders fractional spin unavoidable: the Maxwell term is dangerously irrelevant. 

The above argument assumes a particular order of limits. When we consider Witten's point-splitting regularization, there is also the length scale $a_{\mathrm{f}}$ associated with the point-splitting, as in Eq. \eqref{eq: SL}. If this scale is kept longer than $a_{\mathrm{M}}$ as we take the infrared (IR) limit $a_{\mathrm{M}}\rightarrow0$, then we would obtain $SL[\ell]$ rather than $W[\ell]$. In other words, in this order of limits, it is as if the Maxwell term was never introduced. In the presence of long-ranged interactions, we might think that such an order of limits would be allowed since the Maxwell term is not required to suppress fluctuations of the emergent gauge field  
%One might, in defense of modular invariance, argue that a regularization without Maxwell terms is still possible for theories with long-ranged interactions, as such interactions effectively suppress fluctuations of the emergent gauge field 
 (without long-ranged interactions, the Maxwell term must be included for this purpose). However, to obtain correctly propagating matter at criticality, the particle-vortex duals of these theories \emph{must} have nonvanishing Maxwell terms \cite{Herzog2007}. To see this, notice that the core energy term, 
\be
\mathcal{L}_{\mathrm{core}}=\frac{\varepsilon_c}{2}J_\mu^2(r)\,,
\ee
is a Maxwell term for the emergent gauge field $b$ in the dual theory since particle-vortex duality relates $J_\mu=\epsilon^{\mu\nu\rho}\Delta_\nu b_\rho/2\pi$. Core energy terms can be rewritten as the kinetic terms for the phase fluctuations of the matter fields, and so are crucial for giving rise to the right kinetic terms for the matter fields as we approach criticality. %Taking the inverse point of view, if we start at criticality and deform into the phase, a Maxwell term will inevitably be generated along the renormalization group flow, and it will generically appear in the resulting loop model description. Attempting to tune this term away would not get rid of it in the particle-vortex dual by the above argument. 
Thus, it is not possible to simultaneously eliminate the Maxwell term in both a theory and its particle-vortex dual, and so it is %therefore 
inconsistent to take $a_{\mathrm{M}}\rightarrow0$ before $a_{\mathrm{f}}\rightarrow0$.

\subsection{Fractional Spin and Conformal Field Theories}

The arguments we have presented in this section are well defined in the UV with a specific regulator assumed. Such an analysis amounts to defining a continuum field theory for the loop model. From the discussion above, it is clear that this limit must be subtle given that Witten's and Polyakov's regularizations are not equivalent. Furthermore, as the non-trivial fixed point is approached, the relevant loop configurations become large and fractal-like (reflecting the anomalous dimensions at the fixed point), hence reaching all the way from the UV to the IR. An understanding of these limits is essentially what is needed for a ``derivation'' of the  conjectured web of field theory dualities of Refs. \cite{Seiberg2016,Karch2016}.

The arguments of the previous subsection immediately imply that Polyakov's regularization, in which fractional spin appears, is significantly more natural than Witten's point splitting procedure. We therefore conclude that any loop model description of CFTs of the general form, 
\be
\label{eq: continuum CFT again}
\mathcal{L}_{\mathrm{CFT}}=\mathcal{L}_{\mathrm{matter}}[a]-\frac{1}{4e^2}f^{\mu\nu}\frac{i}{\sqrt{\pd^2}}f_{\mu\nu}+\frac{k}{4\pi}ada\,,
\ee
should include fractional spin. This is because we can always build a loop model by deforming these theories into a phase, and this loop model will generically include fractional spin. Unlike the loop model of Ref. \cite{Fradkin1996}, loop models with fractional spin do not display invariance under periodicity $\mathcal{T}$, so we no longer encounter the issue that periodicity relates theories with different phase diagrams, which should not display duality. Moreover, loop models with fractional spin should yield transport predictions consistent with those of the duality web in Section \ref{subsubsection: conflict with duality web}. This is because Polyakov's duality, Eq. \eqref{eq: Polyakov duality without gauge fields}, uses fractional spin to relate a free massive Dirac fermion to a massive boson with strong short-ranged interactions coupled to a Chern-Simons gauge field at level $\pm1$. Extrapolated to criticality, this duality would simply be the one featured in Eq. \eqref{eq: free fermion duality}, so the transport predictions of this duality would match those of Section \ref{subsubsection: conflict with duality web} (we will explain how to couple Polyakov's duality to background fields in Section \ref{section: Polyakov's duality with gauge fields}).

%and so their partition functions can only display invariance under particle-vortex duality $\mathcal{S}$, not $\mathcal{T}$ nor $\mathcal{T}^2$. This is something of a relief, since we have already seen that fractional spin is, in the language of the loop model, responsible for breaking the discrete spacetime symmetries $\mathbf{P},\mathbf{T},$ and $\mathbf{PH}$, which we na\"{i}vely expect to be broken by the Chern-Simons term in Eq. \eqref{eq: continuum CFT again}. Starting in a phase described by a loop model of the form \eqref{eq: FK model} and approaching criticality, we have already argued that to obtain propagating matter we must include short-ranged interactions in the form of core energies/Maxwell terms, which imply fractional spin. Formally, the partition functions of these theories therefore cannot be invariant under shifts of the statistical angle by $2\pi$.

A more subtle question is whether periodicity somehow survives in any of the correlation functions or critical exponents of the theories  of Eq. \eqref{eq: continuum CFT again}, even though it does not appear in general. This is one way of phrasing the problem of superuniversality of quantum Hall plateau transitions. %(for example, in the fractal properties discussed earlier). 
%Many past analyses have, after all, argued that it is possible to neglect fractional spin because it only affects IR properties of a phase in a simple way through terms that depend only on background fields. However, if the partition function in a phase is not left invariant under periodicity, then we cannot expect the critical theory to be. %Moreover, the critical theory knows about UV properties of the phase: some otherwise massive field is becoming massless. 
On general grounds, because we lack a duality relation between theories related by periodicity, there is no reason to expect the theories of Eq. \eqref{eq: continuum CFT again} to have observables which are invariant under periodicity. For example, by the arguments of Section \ref{subsubsection: conflict with duality web}, we do not expect that DC transport in these theories has a simple transformation law under periodicity. However, there is some reason to be optimistic about critical exponents: recently it has been argued using non-Abelian bosonization dualities that certain theories related by periodicity share correlation length exponents \cite{Hui2017}.

We now return to the question of whether there is any CFT for which the model of Ref. \cite{Fradkin1996}, with full or partial modular invariance, is a good lattice regularization. The answer seems to be negative. As argued above, fractional spin is quite generic, and it prevents us from using generic theories of Chern-Simons gauge fields coupled to gapless matter. However, perhaps there exists an exotic CFT (either local or nonlocal) which can realize periodicity as a symmetry of the partition function along with self-duality. This is an open question. %Such a CFT might manage to retain some topological character, as in Ref. \cite{Aganagic2017}. 

\section{Fractional Spin and a Duality Web of Loop Models}
%\section{Matching Fractional Spin Across Bosonization Dualities}
%\section{Fractional Spin and the Duality Web}
\label{section: duality web}

Having argued that any loop model with hope of describing Chern-Simons theories coupled to matter should include fractional spin, we can ask whether such loop models satisfy the dualities of Refs. \cite{Seiberg2016} and \cite{Karch2016}. %Indeed, we find loop model dualities which parallel the duality web, including its predictions for Abelian theories with general Chern-Simons levels. 
Our strategy will be to use Polyakov's duality, Eq. \eqref{eq: Polyakov duality without gauge fields}, which expresses the partition function of a massive fermion as a bosonic loop model with fractional spin, to derive new dualities. This parallels the philosophy of Refs. \cite{Seiberg2016} and \cite{Karch2016}, which derives the duality web of field theories starting from the assumption of the duality between a gauged Wilson-Fisher boson and a free Dirac fermion. An advantage of working with bosonic loop models is that we never have to work with fermionic matter explicitly. Instead, we derive dualities of the corresponding bosonic loop models. As a result, it is inconvenient to derive boson-boson dualities starting from the seed bosonization duality of Eq. \eqref{eq: general Polyakov duality}. Such dualities are better thought of as following from the Peskin-Halperin-Dasgupta procuedure \cite{Peskin1978,Dasgupta1981} for deriving the particle-vortex duality of lattice loop models. This duality is exact in these models assuming that the statistical interactions between the loops (including fractional spin) can be suitably defined on a lattice \cite{Fradkin1996}.

%Note that we continue to neglect the effects of short-ranged interactions in the loop models beyond the requirement that loops do not cross, so we caution against extrapolating the loop model dualities derived below to criticality. Instead, they should be thought of as a consistency check of the duality web deep in a phase.

%While fractional spin breaks duality under periodicity, it remains to interpret what it means to shift the statistical angle in its presence, as higher spin particles are trivial in 2+1 dimensions. The duality web of Refs. \cite{Seiberg2016,Karch2016} provides a setting in which this is possible. In particular, we find loop models with shifted fractional spin factors to underly Abelian dualities between general bosonic and fermionic Chern-Simons-matter theories that lie in the duality web of Refs. \cite{Seiberg2016,Karch2016}. This follows from the fact that we can check that in a particular phase the loop models corresponding to the dual theories have the same partition function, and so it amounts to an extension of Polyakov's argument \cite{Polyakov1988} to a fuller portion of the duality web. Note that we continue to neglect the effects of short-ranged interactions in the loop models beyond the requirement that loops do not cross, so we caution against extrapolating the loop model dualities derived below to criticality. Instead, they should be thought of as a consistency check on the duality web deep in a phase.

%EF here

\subsection{Coupling Polyakov's Duality to Gauge Fields}
\label{section: Polyakov's duality with gauge fields}

In order to obtain new loop model dualities from Polyakov's duality, Eq. \eqref{eq: Polyakov duality without gauge fields}, we must couple the loop variables to a gauge field $A$, which here we will take to be a background field satisfying the Dirac quantization condition,
\be
\label{eq: flux quantization}
\int_{S^{2}}\frac{dA}{2\pi}\in\mathbb{Z}\,,
\ee
for any $S^2$ submanifold of the spacetime. In theories of a single Dirac fermion, coupling to gauge fields leads to the parity anomaly, so we should expect the loop model partition function Eq. \eqref{eq: Polyakov duality without gauge fields} to also exhibit the parity anomaly. To see how this works, we start with a theory of massive scalar bosons coupled to a Chern-Simons gauge field at level $+1$,\footnote{Throughout this section, we use a metric with the Minkowski signature.} 
\be
|D_a\phi|^2-m_0^2|\phi|^2-|\phi|^4+\frac{1}{4\pi}ada\,.
\ee
This is the bosonic theory in Polyakov's duality, and its partition function can be rewritten as the loop model on the right hand side of Eq. \eqref{eq: Polyakov duality without gauge fields}. $m_0$ is related to the mass $m$ in that equation, but it is not exactly equal to it \cite{SHAJI1990}. Notice that we work in the symmetric (insulating) phase of the theory where the global $U(1)$ symmetry is unbroken, so that $a$ is not Higgsed. We couple this theory to $A$ as follows, 
\be
\label{eq: massive boson}
|D_a\phi|^2-m_0^2|\phi|^2-|\phi|^4+\frac{1}{4\pi}ada+\frac{1}{2\pi}adA\,.
\ee
%We are permitted to do this because it was already shown by Polyakov that the fractional spin imparted on $\phi$ by $a$ leads to the loop model partition function of a Dirac fermion, so 
Coupling this theory to a gauge field should be the same as coupling the Dirac fermion to a gauge field. The theory in Eq. \eqref{eq: massive boson} can be rewritten in a more useful form by shifting $a\rightarrow a+A$,
\be
|D_{a-A}\phi|^2-m_0^2|\phi|^2-|\phi|^4+\frac{1}{4\pi}ada-\frac{1}{4\pi}AdA\,.
\ee
This theory is anomaly free and gauge invariant by construction, so the Dirac fermion it describes should have the right parity anomaly term to enforce gauge invarance. The loop model partition function for this theory has the same form as that with $A=0$, except now $J$ couples to $A$, and we have the background Chern-Simons term
\be
\label{eq:Zboson}
Z[A]=\int\mathcal{D}J\mathcal{D}a\,\delta(\pd_\mu J^\mu)\,e^{-|m|L[J]+iS[J,a,A]}\,,
\ee
where
\be
S[J,a,A]=\int d^3x\left[J(a-A)+\frac{1}{4\pi}ada-\frac{1}{4\pi}AdA\right]\,,
\ee
and we suppress all contractions of spacetime indices in the action. Please note that in Eq. \eqref{eq:Zboson}, as in previous sections, short-ranged interactions are not made explicit. The manipulations that follow in the context of Chern-Simons theory are only consistent if  the  bosons have (strong) short-ranged repulsive interactions. This is also natural since for $D<4$ spacetime dimensions the free massless scalar field fixed point is essentially inaccessible. 

As we know well now, integrating out $a$ results in a nonlocal linking number term for $J$
\be
\label{eq: -pi loop model}
%S_-[J,A]=
-\pi\,\Phi[J]+\int d^3x\left[JA-\frac{1}{4\pi}AdA\right]\,,
\ee
where we have changed variables $J\rightarrow-J$ since the partition function does not depend on the overall sign of $J$. For $A=0$, we recover Eq. \eqref{eq: Polyakov duality without gauge fields} with $\theta=-\pi$. Since the $J$ variables are gapped and bosonic, the response of this theory is determined solely by the background Chern-Simons term, which gives a Hall conductivity $\sigma_{xy}=-\frac{1}{4\pi}$, precisely what we would expect from a massive, properly regulated Dirac fermion, which has a parity anomaly term\footnote{For a derivation using $\zeta$ function regularization see Ref. \cite{Gamboa-1996}.}
\be
\label{Eq: massive Dirac action}
\bar{\Psi}(i\slashed{D}_A-M)\Psi-\frac{1}{8\pi}AdA\,,
\ee  
with $M<0$. Indeed, this is what one finds by following Polyakov's logic starting with Eq. \eqref{eq: -pi loop model}. This identification already suggests that the sign of the linking number term is identified with the sign of the mass of the fermion in the phase. See Appendix \ref{appendix: Polyakov} for a more explicit justification of this statement. Again, $M$ is related, but not equal, to $m$. 

\begin{comment}
What about the theory with statistical angle $\theta=+\pi$? This is obtained by acting time-reversal $\mathbf{T}$ on the theory of Eq. \eqref{eq: massive boson}, which flips the signs of the Chern-Simons and BF terms. The resulting loop model action is
\be
%S_+[J,A]=
+\pi\,\Phi[J]+\int d^3x\left[JA+\frac{1}{4\pi}AdA\right]\,,
\ee
consistent with the Dirac fermion theory,
\be
\bar{\Psi}(i\slashed{D}_A+M)\Psi+\frac{1}{8\pi}AdA\,.
\ee
\end{comment}
Thus, we find that the properly regulated loop model partition function for a Dirac fermion in its $\mathbf{T}$-broken (integer quantum Hall) phase is
%\begin{widetext}
\be
%\label{eq: Polyakov's duality}
\label{eq: Zfermion M < 0}
Z_{\mathrm{fermion}}[A;M<0]\,e^{ -i\operatorname{CS}[A]/2}=\int\mathcal{D}J\,\delta(\pd_\mu J^\mu)\,e^{-|m|L[J]+iS_{\mathrm{fermion}}[J,A;M<0]}e^{-i\operatorname{CS}[A]/2}\,,
\ee
%\end{widetext}
where
\begin{comment}
\begin{align}
\label{eq: Sfermion}
S_{\mathrm{fermion}}[J,A;M]=&\int d^3x\,JA\\&+\,\sgn(M)\left(\pi\Phi[J]+\frac{1}{2}\operatorname{CS}[A]\right)\,,\nonumber
\end{align}
\end{comment}
\begin{align}
\label{eq: Sfermion M < 0}
S_{\mathrm{fermion}}[J,A;M<0]=&\int d^3x\,JA-\,\left(\pi\Phi[J]+\frac{1}{2}\operatorname{CS}[A]\right)\,,
\end{align}
and we define
\be
\operatorname{CS}[A]=\int d^3x\,\frac{1}{4\pi}AdA\,.
\ee
Our reason for factoring out a $\operatorname{CS}[A]/2$ term in Eq. \eqref{eq: Zfermion M < 0} is to isolate the effect of the parity anomaly, which can be thought of as arising from a heavy fermion doubler (or regulator). %Eq. \eqref{eq: Polyakov's duality} demonstrates how we can identify the sign of the linking number term with the sign of the fermion mass in the phase. See Appendix \ref{appendix: Polyakov} for a more explicit justification of this. 
In practice, Eq. \eqref{eq: Zfermion M < 0} tells us how to write the loop model partition function of a Dirac fermion in its $\mathbf{T}$-broken phase, where the low energy effective action is $-\operatorname{CS}[A]$. 

Having completed our analysis for the unbroken phase of Eq. \eqref{eq: massive boson}% and its time reversed conjugate
, how do we write the loop model partition function in the broken symmetry (superfluid) phase of Eq. \eqref{eq: massive boson}? In the Dirac fermion picture, this should be the $\mathbf{T}$-symmetric (trivial insulator) phase, obtained from Eq. \eqref{Eq: massive Dirac action} with $M>0$.
%\be
%i\bar{\Psi}\slashed{D}_A\Psi+M\bar{\Psi}\Psi-\frac{1}{8\pi}AdA\,.
%\ee
Instead of modeling the broken symmetry phase of the $\phi$ variables directly, we use bosonic particle-vortex duality \cite{Peskin1978,Dasgupta1981} to exactly write the loop model partition function in this phase as one describing the symmetric (insulator) phase of vortex variables $\tilde{\phi}$ (with corresponding loop variables $\tilde{J}$). Recall that this changes the dependence on the background fields and inverts the sign of the linking number term, as we saw with the loop model in Section \ref{section: FK model} and demonstrate in Appendix \ref{appendix: self-duality}. This leads to a loop model with statistical angle $\theta=+\pi$. The loop model partition function in the $\mathbf{T}$-symmetric phase is therefore
\be
\tilde{Z}[A]=\int\mathcal{D}\tilde{J}\mathcal{D}b\,\delta(\pd_\mu \tilde{J}^\mu)\,e^{-|m|L[\tilde{J}\,]+i\tilde{S}[\tilde{J},b,A]}\,,
\ee
where
\be
\tilde{S}[\tilde{J},b,A]=\int d^3x\left[\tilde{J}(b-A)-\frac{1}{4\pi}bdb\right]\,.
\ee
Integrating out $b$ and changing variables $\tilde{J}\rightarrow-\tilde{J}$ yields the action
\be
+\pi\Phi[\tilde{J}\,]+\int d^3x\,\tilde{J}A\,,
\ee
consistent with a Hall conductivity $\sigma_{xy}=0$, as we would expect from Eq. \eqref{Eq: massive Dirac action} with $M>0$. Following through with Polyakov's argument from here allows us to write
%\begin{widetext}
\be
%\label{eq: Polyakov's duality}
\label{eq: Zfermion M > 0}
Z_{\mathrm{fermion}}[A;M>0]\,e^{ -i\operatorname{CS}[A]/2}=\int\mathcal{D}J\,\delta(\pd_\mu J^\mu)\,e^{-|m|L[J]+iS_{\mathrm{fermion}}[J,A;M>0]}e^{-i\operatorname{CS}[A]/2}\,,
\ee
%\end{widetext}
where
\begin{align}
S_{\mathrm{fermion}}[J,A;M>0]=&\int d^3x\,JA+\,\left(\pi\Phi[J]+\frac{1}{2}\operatorname{CS}[A]\right)\,,
\end{align}

Bringing everything together, the loop model partition function of a free Dirac fermion with Lagrangian,
\be
\bar{\Psi}(i\slashed{D}_A-M)\Psi-\frac{1}{8\pi}AdA\,,
\ee
having fixed the sign of the parity anomaly term, is
%\begin{widetext}
\bea
\label{eq: general Polyakov duality}
Z_{\mathrm{fermion}}[A;M]e^{- i\operatorname{CS}[A]/2}&=&\det[i\slashed{D}_A-M]e^{-i\operatorname{CS}[A]/2}\\
&=&\int\mathcal{D}J\,\delta(\pd_\mu J^\mu)\exp\left(-|m|L[J]+iS_{\mathrm{fermion}}[J,A;M]-\frac{i}{2}\operatorname{CS}[A]\right).\nonumber
\eea
%\end{widetext}
where the loop model action $S_{\mathrm{fermion}}$ for general $M$ is
\begin{align}
\label{eq: Sfermion}
S_{\mathrm{fermion}}[J,A;M]=&\int d^3x\,JA+\,\sgn(M)\left(\pi\Phi[J]+\frac{1}{2}\operatorname{CS}[A]\right)\,,
\end{align}
Here, too, we have left implicit the necessary interactions between the loops.
Eq. \eqref{eq: general Polyakov duality} is the main result of this subsection. 

In field theory language, we have derived the duality of a massive free Dirac fermion
\begin{equation}
\bar{\Psi}(i\slashed{D}_A-M)\Psi-\frac{1}{8\pi}AdA
\end{equation}
to a gauged Wilson-Fisher scalar with a mass term,
\begin{equation}
|D_a\phi|^2-r|\phi|^2-|\phi|^4+\frac{1}{4\pi}ada+\frac{1}{2\pi}adA\,,
\label{eq:LB-dual}
\end{equation}
starting from the loop model representation of the scalar theory. For $r>0$, the scalar theory is in its symmetric phase, which we showed can be related to the $\mathbf{T}$-broken, $M<0$ phase of the fermionic theory, which has $\sigma_{xy}=-1/(4\pi)$. Conversely, the $\mathbf{T}$-symmetric, $M<0$ phase of the fermionic theory is dual to the symmetry broken phase of the scalar theory, where $r<0$. In this phase, $\sigma_{xy}=0$.
%\begin{equation}
%\mathcal{L}_B=|D_{a}\phi|^2-m_0^2|\phi|^2-|\phi|^4-\frac{1}{4\pi}ada -\frac{1}{2\pi} adA-\frac{1}{4\pi} AdA
%\end{equation}

We can also consider acting time reversal $\mathbf{T}$ on the duality \eqref{eq: general Polyakov duality}. This flips the signs of the Chern-Simons and BF terms, as well as the fermion mass $M$, i.e. this duality corresponds to Eq. \eqref{eq: general Polyakov duality} with $M\rightarrow-M$ and a parity anomaly term with positive sign. In this case, the $\mathbf{T}$-broken phase now has Hall conductivity $\sigma_{xy}=+\frac{1}{4\pi}$. 

%Indeed, in the absence of background fields, particle-vortex duality of the bosonic loop models can be identified with time reversal in their fermionic duals. 

%Thus, as we may have expected, the properly regulated world line partition function for a Dirac fermion of mass $m$ is
%\bea
%Z_{\mathrm{fermion}}[A;m]e^{\pm i\operatorname{CS}[A]/2}&=&\int\mathcal{D}J\,\delta(\pd_\mu J^\mu)e^{-|m|L[J]+iS_{\mathrm{fermion}}[J,A;m]}e^{\pm i\operatorname{CS}[A]/2}\\
%S_{\mathrm{fermion}}[J,A;m]&=&\int d^3x\,JA+\sgn(m)\left(\pi\,\Phi[J]+\frac{1}{2}\operatorname{CS}[A]\right)
%\eea

\subsection{A Duality Web of Loop Models}

%Using Polyakov's duality \eqref{eq: Polyakov's duality}, we can derive a web of dualities which parallels that of Refs. \cite{Seiberg2016,Karch2016}. Our approach will be to start with a proposed duality and check that the corresponding loop model partition functions in a particular phase are equal. An advantage of working with loop models is that we never have to work with fermionic matter explicitly. Instead, we derive dualities of the corresponding bosonic loop models. As a result, it is inconvenient to derive boson-boson dualities starting from the seed duality \eqref{eq: Polyakov's duality}, as is done in \cite{Seiberg2016,Karch2016}. Instead, such dualities are better thought of as following from the Peskin-Halperin-Dasgupta procuedure \cite{Peskin1978,Dasgupta1981} for deriving the particle-vortex duality of lattice loop models. This duality exact in these models assuming that the statistical interactions between the loops (including fractional spin) can be suitably defined on a lattice.

\subsubsection{Fermionic Particle-Vortex Duality}

Equipped with the loop model partition function for a Dirac fermion coupled to a gauge field, we now proceed to derive new loop model dualities. We start by deriving a loop model version of the duality between a free Dirac fermion and 2+1 dimensional quantum electrodynamics (QED$_3$) \cite{Son2015,Wang2015,Metlitski2016}\footnote{As usual, we will only explicitly include the leading relevant operators and suppress irrelevant operators. In particular, Maxwell terms $-\frac{1}{4g_\mathrm{M}^2}f^2$ for the gauge fields are always implicitly included. Thus, the loop model dualities derived here are only meant to hold at energies $E<<g_\mathrm{M}^2$. },
\begin{align}
\label{eq: fermion particle-vortex}
i\bar{\Psi}\slashed{D}_A\Psi-\frac{1}{8\pi}AdA
&\longleftrightarrow
i\bar{\psi}\slashed{D}_a\psi-\frac{1}{4\pi}adA-\frac{1}{8\pi}AdA\,.
\end{align}
Here, it will be more convenient to consider the version of this duality with properly quantized coefficients of Chern-Simons and BF terms in the strongly interacting theory \cite{Seiberg2016},
\begin{align}
i\bar{\Psi}\slashed{D}_A\Psi-\frac{1}{8\pi}AdA
&\longleftrightarrow
i\bar{\psi}\slashed{D}_a\psi+\frac{1}{8\pi}ada-\frac{1}{2\pi}adb+\frac{2}{4\pi}bdb-\frac{1}{2\pi}bdA\,,
\end{align}
where Eq. \eqref{eq: fermion particle-vortex} can be recovered by integrating out the auxiliary gauge field $b$, which comes at the cost of violating flux quantization, Eq. \eqref{eq: flux quantization}. Note that the signs of the $\frac{1}{8\pi}ada$ and $\frac{1}{8\pi}AdA$ terms, which can be thought to arise from heavy fermion doublers coupled to $a$ and $A$ respectively, need not match across this duality.

To obtain loop models, we add a mass term $-M\bar{\Psi}\Psi$, $M<0$, to the free theory and a mass term $-M'\bar{\psi}\psi$, $M'> 0$, to QED$_3$ so that both theories are in their %integer quantum Hall phase (i.e. the phase with broken time-reversal invariance)
$\mathbf{T}$-broken phase. The partition function of the free theory is Eq. \eqref{eq: general Polyakov duality} with $M<0$. Similarly, we can obtain a loop model analogue of QED$_3$ by acting $\mathbf{T}$ on Eq. \eqref{eq: general Polyakov duality}, plugging in $M'>0$, gauging $A\rightarrow a$, and adding the correct couplings to $b$
%\begin{widetext}
\be
Z_{\mathrm{QED}_3}[A;M' > 0]\,e^{-i\operatorname{CS}[A]/2}=\int\mathcal{D}J\mathcal{D}a\mathcal{D}b\,\delta(\pd_\mu J^\mu)e^{-|m|L[J]+iS_{\mathrm{QED}_3}[J,a,b,A;M' < 0]}\,,
\ee
where
\be
S_{\mathrm{QED}_3}[J,a,b,A; M' > 0]=\pi\,\Phi[J]+\int d^3x\left[Ja+\frac{1}{4\pi}ada-\frac{1}{2\pi}adb+\frac{2}{4\pi}bdb-\frac{1}{2\pi}bdA\right]\,.
\ee
We can integrate out $a$ without violating flux quantization to obtain
\begin{align}
S_{\mathrm{eff}}=&\pi\,\Phi[J]+\int d^3x\left[-\pi\left(J-\frac{db}{2\pi}\right)d^{-1}\left(J-\frac{db}{2\pi}\right)+\frac{2}{4\pi}bdb-\frac{1}{2\pi}bdA\right]\nonumber\\
=&\int d^3x\left[Jb+\frac{1}{4\pi}bdb-\frac{1}{2\pi}bdA\right]\,,
\end{align}
%\end{widetext}
where we have used the fact that $\Phi[J]=\int Jd^{-1}J$. Integrating out $b$ gives the action of Eq. \eqref{eq: general Polyakov duality} with $M<0$, so we obtain the loop model duality 
\be
Z_{\mathrm{fermion}}[A;M<0]=Z_{\mathrm{QED}_3}[A;M'>0]\,.
\ee
If we instead work with the trivial insulating phase, similar manipulations lead to
\be
\label{eq: fermion particle-vortex, insulating phase}
Z_{\mathrm{fermion}}[A;M>0]=Z_{\mathrm{QED}_3}[A;M'<0]\,.
\ee

The interpretation of these loop model dualities as fermionic particle-vortex dualities is immediate. First, since the sign of the linking number term is the same as the sign of the mass of the fermion, we recover the mapping of mass operators $\bar{\Psi}\Psi\longleftrightarrow-\bar{\psi}\psi$. What's more, if we violate flux quantization by integrating out $b$, we recover the matter-flux mapping:
\be
\label{eq: fermion particle-vortex map}
J_\Psi\longleftrightarrow\frac{1}{4\pi}da\,,
\ee
where $J^\mu_\Psi=\bar{\Psi}\gamma^\mu\Psi$ is the global $U(1)$ current of the free fermion. This may seem odd since we never actually changed variables in deriving these dualities. However, we were never working with fermionic variables to begin with, but \emph{bosonic} ones. Thus, the loop variables above should not be interpreted as the currents of the free fermion. Instead, Polyakov's duality \eqref{eq: general Polyakov duality} makes clear that since $A$ couples to $J_\Psi$ in $S_{\mathrm{fermion}}$, correlation functions of $J_\Psi$ are generated by derivatives of 
\be
F_{J_\Psi}[A;M]=\log Z_{\mathrm{fermion}}[A;M]\,,
\ee
where we have subtracted off the $\pm\frac{1}{2}\mathrm{CS}[A]$ parity anomaly term, as it does not contribute to the correlation functions of $J_\Psi$. Since $A$ couples as $\frac{1}{4\pi}Ada-\frac{1}{8\pi}AdA$ in $S_{\mathrm{QED_3}}$ after $b$ is integrated out, subtracting off the same parity anomaly term in the QED$_3$ theory implies the mapping of Eq. \eqref{eq: fermion particle-vortex map}. 

%\subsubsection{Abelian Bosonization Dualities from Matching Fractional Spin}
\subsubsection{General Abelian Bosonization Dualities}

%Equipped with the loop model partition function for a Dirac fermion coupled to a gauge field, we now proceed to derive more general loop model bosonization dualities. 
We now consider more general boson-fermion dualities. The field theory duality web \cite{Seiberg2016,Karch2016} can be used to relate a theory of a Wilson-Fisher scalar coupled to a Chern-Simons gauge field at level $k_\phi\in\mathbb{Z}$,
\be
\label{eq: general CS-scalar}
|D_a\phi|^2-|\phi|^4+\frac{k_\phi}{4\pi}ada+\frac{1}{2\pi}adA\,,
\ee
to a dual theory of Dirac fermions coupled to a Chern-Simons gauge field. To do this, we invoke the duality between a Wilson-Fisher boson and a Dirac fermion coupled to a Chern-Simons gauge field at level 1/2 
\be
|D_{A}\phi|^2-|\phi|^4\longleftrightarrow i\bar{\psi}\slashed{D}_b\psi+\frac{1}{8\pi}bdb+\frac{1}{2\pi}bdA+\frac{1}{4\pi}AdA\,.
\ee
Plugging this result into Eq. \eqref{eq: general CS-scalar}, we obtain a duality
\begin{align}
\label{eq: general bosonization duality}
|D_a\phi|^2-|\phi|^4+\frac{k_\phi}{4\pi}ada+\frac{1}{2\pi}adA
&\longleftrightarrow
i\bar\psi\slashed{D}_b\psi+\frac{1}{8\pi}bdb+\frac{1}{2\pi}ad(b+A)+\frac{k_\phi+1}{4\pi}ada\,.
\end{align}
Integrating out the gauge field $a$ on the fermionic side, would run into conflict with flux quantization Eq. \eqref{eq: flux quantization} and gauge invariance (if the theory is defined purely in 2+1 dimensions). However, continuing in spite of this, one would obtain a duality between bosons coupled to a Chern-Simons gauge field at level $k_\phi$ and fermions coupled to a Chern-Simons gauge field at level \cite{Mross2017}
\be
\label{eq: kpsi}
k_\psi=\frac{1}{2}\frac{k_\phi-1}{k_\phi+1}\,.
\ee
This relation can be generalized to the self-dual theories which occupied our attention for much of this work, Eq. \eqref{eq: continuum CFT again}, which, in addition to Chern-Simons terms, have marginally long-ranged interactions. The introduction of such long-ranged interactions can be accommodated by replacing $k_\phi$ and $k_\psi$ with $\tau_{\phi}=k_{\phi}+i\frac{2\pi}{e_{\phi}^2}$ and $\tau_{\psi}=2k_\psi+i\frac{4\pi}{e^2_\psi}$ respectively,
\be
\tau_\psi=\frac{\tau_\phi-1}{\tau_\phi+1}\,.
\ee
For clarity, in this section we will only explicitly consider the limit $e^2_{\phi,\psi}\rightarrow\infty$. Our results can be readily generalized away from this limit by replacing $k$'s with $\tau$'s.

We can easily check that the theories on either side of the duality Eq. \eqref{eq: general bosonization duality} have the same phase diagram. Adding $+m^2|\phi|^2$ to the scalar theory Higgses out the emergent gauge field $a$ and leaves the theory in a trivial insulating phase. Similarly, adding $-M\bar{\psi}\psi$, with $M<0$, to the fermion theory and integrating out $b$ also Higgses out $a$ (the parity anomaly term being cancelled), leading to the same phase. Conversely, adding $-m^2|\phi|^2$ to the scalar theory leads to a topological quantum field theory of the form
\be
\label{eq: tqft phase}
\frac{k_\phi}{4\pi}ada+\frac{1}{2\pi}adA\,,
\ee
which can also be obtained on the fermionic side by adding $-M\bar{\psi}\psi$ with $M>0$. Integrating out $\psi$ adds a parity anomaly term to the action,
\be
\label{eq: phase for fermion}
\frac{1}{4\pi}bdb+\frac{1}{2\pi}ad(b+A)+\frac{k_\phi+1}{4\pi}ada\,.
\ee
Integrating out $b$ leads to Eq. \eqref{eq: tqft phase}. Our interest will be in this phase: our goal will be to show that in this phase the loop model partition function of massive fermions coupled to $b$ is the same of that of the massive bosons coupled to $a$. %As with the fermion-fermion duality derived in the previous subsection, an analogous argument can be made in the topologically trivial phase by working with the particle-vortex dual of Eq. \eqref{eq: general CS-scalar}.

The world line partition function for the gapped bosons in the phase described by Eq. \eqref{eq: tqft phase} is (turning off background fields) 
\begin{equation}
Z_{\mathrm{boson}}[k_\phi]=\int\mathcal{D}J\mathcal{D}a\,\delta(\pd_\mu J^\mu)\,e^{-|m|L[J]+iS_{\mathrm{boson}}[J,a;k_\phi]},
\end{equation}
where we used the definition
\begin{equation}
S_{\mathrm{boson}}[J,a;k_\phi]=\int d^3x\left[J a+\frac{k_\phi}{4\pi}ada\right]\,.
\end{equation}
%where $L[J]$ is the sum of the lengths of the loops in the configuration $J$ and we have suppressed indices in the action. 
Integrating out $a$ yields an effective action 
\be
\label{eq: bosonic effective action}
S_{\mathrm{eff}}[J]=-\frac{\pi}{k_\phi} \Phi[J]\,.
\ee
%where $\Phi[J]=-\int d^3xJd^{-1}J$ is the linking number of the configuration $J$ defined in Eq. \eqref{eq: linking number}, which includes the full writhes of all of the constituent loops of $J$ and therefore is not a topological invariant. 
%Constructing the world line path integral for the fermion is a bit more subtle and is obtained most clearly in a supersymmetric formalism \cite{,PolyakovLesHouches1988,Polyakov:1987ez,Grundberg1990}.
%The world line path integral for the fermion can be written down using Polyakov's original result. We write the partition function for a massive Dirac fermion coupled to a background gauge field $A$ as a bosonic world line path integral with linking number term $\pm i\pi\Phi[J]$ to enforce spin and statistics
%\be
%\label{eq: Polyakov's duality}
%Z_{\mathrm{fermion}}[A]=\int\mathcal{D}J\,\delta(\pd_\mu J^\mu)\, e^{-|m|L[J]}e^{i\left[\pm \pi\Phi[J]+\int d^3x\,\left(JA\mp \frac{1}{4\pi}AdA\right)\right]}
%\ee
%where the background Chern-Simons term is a consequence of the parity anomaly, and its relative sign is fixed to match responses on either side of Polyakov's duality. 

The loop model partition function for the fermion can be written down by acting with $\mathbf{T}$ and gauging the background field $A\rightarrow b$ in Eq. \eqref{eq: general Polyakov duality}. %, where we choose the case where the level is $+1$ to be consistent with Eq. \eqref{eq: phase for fermion}. %The path integral for the massive Dirac fermions coupled to a dynamical gauge field $b$ at level $\pm1$, then, is obtained simply by gauging $A\rightarrow b$ in the above equation. 
The world line partition function for fermions in the phase \eqref{eq: phase for fermion} is therefore
\begin{align}
Z_{\mathrm{fermion}}[k_\psi,M>0]
&=\int\mathcal{D}J\mathcal{D}a\mathcal{D}b\,\delta(\pd_\mu J^\mu)\,e^{-|m|L[J]+iS_{\mathrm{fermion}}[J,b,a;k_\psi,M<0]}\,,
\end{align}
where $k_\psi$ is the level of the gauge field $b$ coupled to the fermion with the auxiliary gauge field $a$ integrated out\footnote{Note that we defined $k_\psi$ as the level of $b$ expected at the critical point, so it does not include the extra parity anomaly term acquired by gapping $\psi$.}, Eq. \eqref{eq: kpsi}, and we define
%\begin{widetext}
\be
S_{\mathrm{fermion}}[J,b,a;k_\psi,M>0]=\pi\, \Phi[J]+\int d^3x\left[J b+\frac{1}{4\pi}bdb+\frac{1}{2\pi}adb+\frac{k_\phi+1}{4\pi}ada\right]\,.
\ee
%\end{widetext}
We now have two options. The first is to integrate out $a$, but that would violate flux quantization, Eq. \eqref{eq: flux quantization}. Instead, we integrate out $b$ first. Its equation of motion is
\be
J+\frac{da}{2\pi}=-\frac{db}{2\pi}\,.
\ee
Charge quantization of the bosonic $J$ variables implies that this equation is consistent with flux quantization. Integrating out $b$ therefore gives
%\begin{widetext}
\begin{align}
S_{\mathrm{eff}}[J,a]&=\pi\, \Phi[J]+\int d^3x\left[-\pi\left(J+\frac{da}{2\pi}\right)d^{-1}\left(J+\frac{da}{2\pi}\right)+\frac{k_\phi+1}{4\pi}ada\right]\nonumber \\
&=\pi\, \Phi[J]-\pi\, \Phi[J]+\int d^3x\left[-Ja-\frac{1}{4\pi}ada+\frac{k_\phi+1}{4\pi}ada\right]\nonumber \\
&=\int d^3x\left[-Ja+\frac{k_\phi}{4\pi}ada\right]\,,
\end{align}
%\end{widetext}
where we have used the definition of $\Phi[J]$ in passing to the second line. The path integral does not depend on the sign of $J$ here, so integrating out $a$ yields the same answer as in the bosonic case, Eq. \eqref{eq: bosonic effective action}. 

We therefore find an equality of the loop model partition functions
\be
Z_{\mathrm{boson}}[k_\phi]=Z_{\mathrm{fermion}}[k_\psi,M>0]\,.
\ee
This is the general three dimensional bosonization identity for loop models. The same analysis can be carried out in the superfluid phase of the theory in Eq. \eqref{eq: general CS-scalar}, which is the insulating phase of its particle-vortex dual, in which the Chern-Simons gauge field has ``level'' $-1/k_\phi$ (again a statement about the theory obtained after violating flux quantization and integrating out auxiliary gauge fields). Denoting the partition function of this theory as $Z_{\mathrm{boson}}[-1/k_\phi]$, we indeed find
\be 
Z_{\mathrm{boson}}[-1/k_\phi]=Z_{\mathrm{fermion}}[k_\psi,M<0]\,.
\ee
Starting from Polyakov's loop model duality \eqref{eq: general Polyakov duality}, we have thus derived a loop model version of the general CFT duality of Eq. \eqref{eq: general bosonization duality} by matching fractional spin factors! %We reiterate that this does not constitute a proof, though, since we did not adequately account for short-ranged interactions, as we discussed in Section \ref{section: fractional spin}.

This derivation also provides an answer to the question of what it means to have fractional spin different from $0$ or $1/2$, a vexing issue since Polyakov's argument first appeared. From the perspective of the duality of Eq. \eqref{eq: general bosonization duality}, it is incorrect to think of theories of world lines with general fractional spin as theories of free particles with a strange spin. Rather, one should think of the theories on either side of the duality as strongly interacting Chern-Simons theories coupled to matter. We further note that these theories are also thought to be dual to non-Abelian Chern-Simons-matter theories \cite{Aharony2016}, but it is not clear to us how to construct an explicit loop model description of these theories. This may be an interesting direction for future work.

\section{Discussion}
\label{section: discussion}

%\section{Discussion}
%\label{section: discussion}

In this article we have shown that, upon introducing fractional spin, 2+1 dimensional loop models with statistical and long-ranged interactions of Eq. \eqref{eq: FK model} are not invariant under shifts of the statistical angle, despite remaining self-dual under particle-vortex duality. This means that, while PSL$(2,\mathbb{Z})$ still has a natural action on these theories, only the $\mathcal{S}$ transformation should be taken as a good duality transformation. It also means that the superuniversal transport properties of the loop models in Ref. \cite{Fradkin1996} do not have an analogue in Chern-Simons theories coupled to gapless matter, which we argued must include fractional spin.

By introducing fractional spin into the loop models of Ref. \cite{Fradkin1996}, we were led to develop simple loop model versions of various members of the web of 2+1 dimensional field theory dualities \cite{Seiberg2016,Karch2016}, starting from a seed duality relating the partition function of a massive Dirac fermion to a bosonic loop model with a fractional spin term. This makes clear the consistency of relativistic loop model dualities with the duality web of conformal field theories. It also should be considered a nontrivial check of these dualities. 

We emphasize that the duality of the loop models suggests that these theories may have a critical point, which should presumably be a relativistic CFT. Proving this statement requires solving the loop model and finding its continuum limit at the phase transition. This has not been done. Our purpose here was to inquire to what extent the critical points of the loop models can be described by a CFT on the duality web. A successful construction of this continuum limit would in fact be a derivation of the duality web.

Because of the simplicity of the loop model dualities presented here, it would be of interest to use loop models to motivate new field theory dualities or derive already proposed dualities which live outside the duality web. However, some difficulties persist. It still remains to carefully implement the short-ranged interactions in the loop model dualities presented here. %and, in particular, to derive the proposed bosonization dualities involving fermions with Gross-Neveu interactions. 
This is a necessary requirement to develop loop model derivations of dualities with multiple flavors of matter fields, which can have different global symmetries depending on the form of the short-ranged interactions. Such dualities have been of interest in the study of deconfined quantum critical points \cite{Wang2017}. It also remains to construct a precise lattice formulation of the loop models presented here, which we defined based on their long distance properties due to the subtleties surrounding placing Chern-Simons theories on a lattice.

\section*{Acknowledgements}
We thank J. Y. Chen, T. Faulkner, T. H. Hansson, S. Kivelson, O. Motrunich, M. Mulligan, S. Raghu, N. Seiberg, S. Shenker, H. Wang, Y. Wang and C. Xu for helpful discussions. H.G. is supported by the National Science Foundation Graduate Research Fellowship Program under Grant No. DGE-1144245. This work was supported in part by National Science Foundation grants DMR 1408713 and DMR 1725401 at the University of Illinois (E.F.).

\begin{appendix}
%\onecolumngrid
\section{Derivation of Self-Duality}
\label{appendix: self-duality}

\subsection{Euclidean Lattice Model}
In this appendix, we derive the self-duality of the Euclidean lattice model of Ref. \cite{Fradkin1996}, 
\be
Z=\sum_{\{J_\mu\}}\delta(\Delta_\mu J^\mu)e^{-S}\,,
\ee
where
\begin{align}
S=&\frac{1}{2}\sum_{r,r'} J^\mu(r) G_{\mu\nu}(r-r')J^\nu(r')+\frac{i}{2}\sum_{r,R}J^\mu(r) K_{\mu\nu}(r,R)J^\nu(R)\nonumber\\
\label{eq: appendix FK model}
&+i\sum_{r,r'}e(r-r')J^\mu(r)A_\mu(r')+\sum_{R,R'}h(R-R')\epsilon^{\mu\nu\rho}J_\mu(R)\Delta_\nu A_\rho(R')\\
&+\frac{1}{2}\sum_{r,r'}A_\mu(r)\Pi^{\mu\nu}(r,r')A_\nu(r')\,.\nonumber
\end{align}
Our conventions and notation are described in Section \ref{section: FK model}. We first consider the case in which the background fields $A_\mu=0$. Following Ref. \cite{Peskin1978}, we invoke the Poisson summation formula to make the formerly integer-valued $J_\mu$ real-valued
\be
Z=\sum_{\{m_\mu\}}\int\mathcal{D}J\,\delta(\Delta_\mu J_\mu)e^{-S[J]+2\pi i\sum_rm_\mu(r) J^\mu(r)}\,,
\ee
where $m_\mu$ is a new integer-valued variable. We can impose the delta function-imposed Gauss' Law $\Delta_\mu J^\mu=0$ by rewriting $J^\mu$ as the curl of an emergent gauge field $a_\mu$
\be
J^\mu(r)=\frac{1}{2\pi}\epsilon^{\mu\nu\lambda}\Delta_\nu a_\lambda(r)\,.
\ee
Plugging this into the action, we obtain
\be
S=\frac{1}{2(2\pi)^2}\sum_{r,r'}a^\mu(r)G_{\mu\nu}(r-r')a^\nu(r')+\frac{i}{2(2\pi)^2}\sum_{r,R}a^\mu(r) K_{\mu\nu}(r-R)a^\nu(R)+i\sum_Ra_\mu(R)\epsilon^{\mu\nu\lambda}\Delta_\nu m_\lambda(R)\,.
\ee
We can now change to vortex loop variables,
\be
\tilde{J}^\mu=\epsilon^{\mu\nu\lambda}\Delta_\nu m_\lambda\,,
\ee
which satisfy their own Gauss' law $\Delta_\mu\tilde{J}^\mu=0$. The partition function in these variables is thus
\be
Z=\sum_{\{\tilde{J_\mu}\}}\int\mathcal{D}a\,\delta(\Delta_\mu\tilde{J}^\mu)e^{-S[\tilde{J},a]}\,.
\ee
We now proceed to integrate out $a$. In the long distance limit, we can use Eqs. \eqref{eq:Gmunu}-\eqref{eq:Kmunu} to write the propagator for $a_\mu$ as
\be
\mathcal{G}^{\mu\nu}(p)=(2\pi)^2\left[\frac{g^2}{g^4+4\theta^2}\frac{1}{|p|}(\delta^{\mu\nu}-p^\mu p^\nu/p^2)+\frac{2\theta}{g^4+4\theta^2}\epsilon^{\mu\nu\lambda}\frac{p_\lambda}{p^2}\right]\,,
\ee
where we have added a gauge fixing term $\frac{1}{2\xi}(\Delta_\mu a^\mu)^2$ in the limit $\xi\rightarrow0$ (Landau gauge). Integrating out $a$ gives the dual loop model action
\be
S_D[\tilde{J}\,]=\frac{1}{2}\sum_{p}\tilde{J}^\mu(-p)\mathcal{G}_{\mu\nu}(p)\tilde{J}^\nu(p)\,.
\ee
In position space, this has the same form as Eq. \eqref{eq: appendix FK model}, but with
\be
\label{eq: transformed couplings}
g^2\mapsto g_D^2=\frac{g^2}{g^4/(2\pi)^2+\theta^2/\pi^2},\,\theta\mapsto\theta_D=-\frac{\theta}{g^4/(2\pi)^2+\theta^2/\pi^2}\,.
\ee
In other words, duality maps $\tau=\frac{\theta}{\pi}+i\frac{g^2}{2\pi}$ as a modular $\mathcal{S}$ transformation
\be
\tau\mapsto-\frac{1}{\tau}\,.
\ee

Now consider the case with background fields turned on, $A_\mu\neq0$. Then $a_\mu$ couples to
\be
i\tilde{J}_\mu+i\frac{e}{2\pi}\,\epsilon_{\mu\nu\lambda}\Delta^\nu A^\lambda+\frac{h}{2\pi}(\Delta^2\delta_{\mu\nu}-\Delta_\mu\Delta_\nu)A^\nu\,.
\ee
When $a_\mu$ is integrated out, this leads to couplings between $\tilde{J}_\mu$ and $A_\mu$ which in momentum space take the form
\bea
&&i\tilde{J}_\rho\mathcal{G}^{\rho\mu}\left(-\frac{e}{2\pi}\,\epsilon_{\mu\nu\lambda}p^\nu A^\lambda+\frac{h}{2\pi}(p^2\delta_{\mu\nu}-p_\mu p_\nu)A^\nu\right)\nonumber\\
&=&\frac{i}{2\pi}\left[\theta_De+|p|g_D^2h\right]\tilde{J}_\rho A^\rho+\frac{i}{2\pi}\left[-\frac{1}{|p|}g_D^2e+\theta_Dh\right]\epsilon^{\mu\nu\rho}\tilde{J}_\mu p_\nu A_\rho\,. 
\eea
So under duality the charges map as
\be
\label{eq: charge map}
e\mapsto\frac{1}{2\pi}\left[\theta_De+|p|g_D^2h\right],\,h\mapsto\frac{1}{2\pi}\left[-\frac{1}{|p|}g_D^2e+\theta_Dh\right]\,,
\ee
meaning that, under particle-vortex duality, electric and magnetic charges are mapped to dyons!

Similarly, the background polarization tensor is shifted under duality. If we define it in momentum space as
\be
\Pi_{\mu\nu}=\Pi_{\mathrm{even}}(p)(\delta_{\mu\nu}-p_\mu p_\nu/p^2)+\Pi_{\mathrm{odd}}(p)\epsilon_{\mu\nu\lambda}\frac{p^\lambda}{|p|}\,,
\ee
then
\bea
\label{eq: polarization tensor map 1}
\Pi_{\mathrm{even}}&\mapsto&\Pi_{\mathrm{even}}+|p|(e^2-h^2p^2)\frac{g_D^2}{(2\pi)^2}-2p^2he\frac{\theta_D}{(2\pi)^2}\,, \\
\label{eq: polarization tensor map 2}
\Pi_{\mathrm{odd}}&\mapsto&\Pi_{\mathrm{odd}}-|p|(e^2-h^2p^2)\frac{\theta_D}{(2\pi)^2}-2p^2eh\frac{g^2}{(2\pi)^2}\,.
\eea

\subsection{Conformal Field Theory}
The above lattice derivation carries over to conformal field theories of the form (now working in Minkowski space)
\bea
\label{eq: appendix CFT}
\mathcal{L}&=&|D_a\phi|^2-|\phi|^4-\frac{1}{4 g_D^2}f_{\mu\nu}\frac{i}{\sqrt{\pd^2}}f^{\mu\nu}+\frac{1}{4\theta_D}ada\\
&&+e(x)J_\mu A^\mu+h(x)JdA+A_\mu(x)\Pi^{\mu\nu}(x,x')A_\nu(x')\,,\nonumber
\eea
where $\phi$ is a complex scalar field at its Wilson-Fisher fixed point (thus the notation $-|\phi|^4$), $D_a=\pd_\mu-ia_\mu$, $f_{\mu\nu}=\pd_\mu a_\nu-\pd_\nu a_\mu$, $A_\mu$ is a background $U(1)$ gauge field, and $J_\mu$ is the global $U(1)$ current $J_\mu=i(\phi^*\pd_\mu\phi-\phi\pd_\mu\phi^*)$. We also use the notation $AdB=\epsilon^{\mu\nu\rho}A_\mu\pd_\nu B_\rho$. For a discussion of the self-duality of this theory with Dirac fermion matter, see Ref. \cite{Mross2017}.

The coupling constants $g_D^2$ and $\theta_D$ are indeed the corresponding quantities in the loop model description \eqref{eq: transformed couplings}. If we deform this model into its symmetric phase (the trivial insulator) with the operator $-m^2|\phi|^2$, we can construct a loop model with Lagrangian
\be
J_\mu a^\mu-\frac{1}{4 g_D^2}f_{\mu\nu}\frac{i}{\sqrt{\pd^2}}f^{\mu\nu}+\frac{1}{4\theta_D}ada\,,
\ee
where we have set $A_\mu=0$ for clarity. Integrating out $a_\mu$ results in a loop model characterized by a modular parameter $\tau=-\left(\frac{\theta_D}{\pi}+i\frac{g_D^2}{2\pi}\right)^{-1}=\frac{\theta}{\pi}+i\frac{g^2}{2\pi}$.  

We can derive the self-duality of the theory \eqref{eq: appendix CFT} by assuming the bosonic particle-vortex duality relating a Wilson-Fisher fixed point to the critical point of the Abelian Higgs model \cite{Peskin1978,Dasgupta1981},
\be
|D_A\phi|-|\phi|^2\longleftrightarrow|D_a\tilde{\phi}|^2-|\tilde{\phi}|^4+\frac{1}{2\pi}adA\,,
\ee
and applying it to the Lagrangian, Eq. \eqref{eq: appendix CFT}. Again turning off background fields for clarity, we obtain
\be
\mathcal{L}\longleftrightarrow|D_b\tilde{\phi}|^2-|\tilde{\phi}|^4+\frac{1}{2\pi}adb-\frac{1}{4 g_D^2}f_{\mu\nu}\frac{i}{\sqrt{\pd^2}}f^{\mu\nu}+\frac{1}{4\theta_D}ada\,.
\ee
Integrating out $a$ in the dual theory gives an action
\be
\tilde{\mathcal{L}}=|D_b\tilde{\phi}|^2-|\tilde{\phi}|^4-\frac{1}{4 g^2}f'_{\mu\nu}\frac{i}{\sqrt{\pd^2}}f'^{\mu\nu}+\frac{1}{4\theta}bdb\,,
\ee
where $f'_{\mu\nu}=\pd_\mu b_\nu-\pd_\nu b_\mu$. $g^2,\theta$ are related to $g_D^2,\theta_D$ by the modular $\mathcal{S}$ transformation \eqref{eq: transformed couplings}. The analogous transformation laws for the source terms Eq. \eqref{eq: charge map} and Eqs. \eqref{eq: polarization tensor map 1}-\eqref{eq: polarization tensor map 2} can be obtained in the same way.

\section{Derivation of Polyakov's Duality}
\label{appendix: Polyakov}

In this appendix, we review Polyakov's argument for the duality \eqref{eq: Polyakov duality without gauge fields}, which relates a theory of \emph{non-intersecting} bosonic loops (i.e. bosonic loops with strong short-ranged repulsion) to a theory of a single free species of Dirac fermions \cite{Polyakov1988}. Some of the finer points of this argument were ironed out in Refs. \cite{SHAJI1990,Alekseev1988,Orland1988,Grundberg1989,Grundberg1990}, to which we point the reader interested in a more detailed analysis. For a review of how to construct world line partition functions from gapped quantum field theories, see Refs. \cite{Polyakov:1987ez,PolyakovLesHouches1988,Fradkin-2013,SHAJI1990}.

We start from the bosonic loop model on the right hand side of Eq. \eqref{eq: Polyakov duality without gauge fields}. The amplitude for a path of length $L$ with tangent vector $\hat{e}(s)$ between two points $x$ and $x'$ is
\be
G(x-x')=\int_0^\infty dL\int \mathcal{D}\hat{e}\,\delta\left(1-|\hat{e}|^2\right)\delta\left(x'-x-\int_0^Lds\,\hat{e}(s)\right)e^{-|m|L\pm i\pi \mathcal{W}[\hat{e}]}\,,
\ee
where $\mathcal{W}[\hat{e}]$ is the Berry phase term in the twist, Eq. \eqref{eq: twist as Berry phase},
\be
\mathcal{W}[\hat{e}]=\frac{1}{2\pi}\int_0^L ds\int_0^1du\,\hat{e}\cdot(\pd_s\hat{e}\times\pd_u\hat{e})\,.
\ee
The momentum space representation is obtained via Fourier transform
\be
G(p)=\int_0^\infty dL\int \mathcal{D}\hat{e}\,\delta\left(1-|\hat{e}|^2\right)e^{-|m|L\pm i\pi \mathcal{W}[\hat{e}]}e^{ip^\mu\int_0^Lds\,\hat{e}_\mu(s)}\,.
\ee
This is none other than the coherent state path integral for a spin-1/2 particle in a magnetic field $b^\mu= \pm2p^\mu$. The equation of motion for $\hat{e}$ is
\be
\pd_s\hat{e}_\mu=\pm2\epsilon_{\mu\nu\lambda}\hat{e}^\nu p^\lambda=i[H,\hat{e}_\mu]\,,
\ee
where $H$ is the Hamiltonian operator. This implies that, since the Hamiltonian for the spin is $H=-\vec{b}\cdot\vec{S}= \mp p^\mu\hat{e}_\mu$, upon quantization $\hat{e}$ should satisfy  commutation relations
\be
[\hat{e}_\mu,\hat{e}_\nu]=2i\epsilon_{\mu\nu\rho}\hat{e}^\rho\,,
\ee
meaning that we can perform the path integral over $\hat{e}$ by identifying it with the Pauli matrices $\hat{e}_\mu\rightarrow\sigma_\mu$. There is one slight problem with the above discussion, however: in reality, the magnetic field is not $2p^\mu$ but $2p^\mu/3$, meaning that this identification involves the appearance of an extra constant factor which can be eliminated by rescaling $m\rightarrow M$ \cite{Orland1988}. Performing the integral over $L$ finally gives the propagator of a free Dirac fermion of mass $M$
\be
G(p)\propto\frac{1}{ip^\mu\sigma_\mu-M}\,,
\ee
where we have written $\pm|M|\equiv M$. One thus expects the partition function \eqref{eq: Polyakov duality without gauge fields} to reproduce the correlation functions of a free Dirac fermion of mass $M$ at sufficiently long distances.

Note that the theory we have discussed here is not coupled to gauge fields. As is well known, when coupled to a gauge field, theories of a single Dirac fermion exhibit the parity anomaly. See Section \ref{section: Polyakov's duality with gauge fields} for a discussion of how the parity anomaly appears in the context of this duality.

\end{appendix}

%\twocolumngrid
\nocite{apsrev41Control}
\bibliography{Self_Duality}{}
\bibliographystyle{apsrev4-1}

\end{document}